\documentclass[aps,showpacs,showkeys,preprintnumbers,nofootinbib]{revtex4}

\usepackage[english]{babel}
\usepackage{graphicx}
\usepackage{color}
\usepackage{amsfonts}
\usepackage{amsmath}
\usepackage{amssymb}

\def \be  {\begin{equation}}
\def \ee  {\end{equation}}
\def \ee  {\end{equation}}
\def \bea {\begin{eqnarray}}
\def \eea {\end{eqnarray}}

\begin{document}

\preprint{ECTP-2016-02}
\preprint{WLCAPP-2016-02}

\title{Particle production and chemical freezeout from the hybrid UrQMD approach at NICA energies}

\author{Abdel Nasser Tawfik}
\email{a.tawfik@eng.mti.edu.eg}
\affiliation{Egyptian Center for Theoretical Physics (ECTP), Modern University for Technology and Information (MTI), 11571 Cairo, Egypt}
\affiliation{World Laboratory for Cosmology and Particle Physics (WLCAPP), Cairo, Egypt}

\author{Loutfy I. Abou-Salem}
\affiliation{Physics Department, Faculty of Science, Benha University, 13518 Benha, Egypt}

\author{Asmaa G. Shalaby}
\affiliation{Physics Department, Faculty of Science, Benha University, 13518 Benha, Egypt}
\affiliation{World Laboratory for Cosmology and Particle Physics (WLCAPP), Cairo, Egypt}

\author{Mahmoud Hanafy}
\affiliation{Physics Department, Faculty of Science, Benha University, 13518 Benha, Egypt}
\affiliation{World Laboratory for Cosmology and Particle Physics (WLCAPP), Cairo, Egypt}

\author{Alexander Sorin}
\affiliation{Bogoliubov Laboratory of Theoretical Physics and Veksler and Baldin Laboratory of High Energy Physics,
Joint Institute for Nuclear Research, 141980 Dubna, Moscow region, Russia}
\affiliation{National Research Nuclear University (MEPhI),  115409 Moscow, Russia}
\affiliation{Dubna International University, 141980, Dubna, Russia}

\author{Oleg Rogachevsky}
\affiliation{Veksler and Baldin Laboratory of High Energy Physics, Joint Institute for Nuclear Research, 141980 Dubna, Moscow region, Russia}

\author{Werner Scheinast}
\affiliation{ Veksler and Baldin Laboratory of High Energy Physics, Joint Institute for Nuclear Research, 141980 Dubna, Moscow region, Russia}

\date{\today}

\begin{abstract}

The energy dependence of various particle ratios is calculated within the Ultra-Relativistic Quantum Molecular Dynamics approach and compared with the hadron resonance gas (HRG) model and measurements  from various experiments, including RHIC-BES, SPS and AGS.  It is found that the UrQMD particle ratios agree well with the experimental results at the RHIC-BES energies. Thus, we have utilized UrQMD in simulating particle ratios at other beam energies  down  to $3~$GeV, which will be accessed at NICA and FAIR future facilities. We observe that  the particle ratios for crossover and first-order phase transition, implemented in the hybrid UrQMD v$3.4$, are nearly indistinguishable, especially at low energies (at large baryon chemical potentials or high density). 

\end{abstract}.

\maketitle


\section{Introduction}

One of the main goals of the heavy-ion experiments is the characterization of strongly interacting matter under extreme conditions of high temperature and density  \cite{r1}. Examining the possible quark-hadron phase transition(s) plays a crucial role in verifying the quantum chromodynamics (QCD), the theory of strong interactions, which predicts that the confined hadrons likely undergo phase transition(s) to partonic matter called quark-gluon plasma (QGP) \cite{r2}. So far, various signatures for the QGP formation have been verified, experimentally \cite{r3}. The statistical-thermal models \cite{r4,r5,r6,r7,r8,r9,r10,r11} are successful approaches explaining - among others - the produced particle yields and their ratios. At chemical equilibrium, it is conjectured that the particle ratios are well described by at least two parameters, the baryon chemical potential ($\mu_{\mathrm{b}})$ and the freezeout temperature ($T_{\mathrm{ch}}$). The chemical freezeout is defined as a stage during the evolution of the high-energy collision at which the inelastic collisions is assumed to disappear and the number of produced particle should be fixed. Experiments at the Schwerionensynchrotron (SIS18) \cite{r12,r13}, the Alternating Gradient Synchrotron (AGS) \cite{r5}, the Super Proton Synchrotron (SPS) \cite{r7}, and the Relativistic Heavy-Ion Collider (RHIC)  \cite{r14,r15,r16,r17} have been successfully reproduced within the statistical-thermal approaches \cite{r4}.

The dependence of both freezeout parameters ($T_{\mathrm{ch}}$ and $\mu_{\mathrm{b}}$) on the nucleon-nucleon center-of-mass energies ($\sqrt{s_{\mathrm{NN}}}$) known as the chemical freezeout boundary looks very similar to the QCD phase-diagram separating confined hadrons from deconfined QGP \cite {r18}.  In lattice QCD simulations \cite {r19,r20}, which are very reliable at $\mu_{\mathrm{b}}/T\leq 1$, i.e., $\sqrt{s_{\mathrm{NN}}}$ greater than top SPS energies, the dependence of $T_{\mathrm{ch}}$ on $\mu_{\mathrm{b}}$ appears very close to the QCD critical line. At larger $\mu_{\mathrm{b}}$ (lower energies), both boundaries become distinguishable \cite{Tawfik:2016jzk}. In this region, lattice QCD simulations suffer from serious numerical difficulties (the so-called sign problem). Thus, we are left with effective models such as statistical-thermal models and QCD-like approaches including linear-sigma and Nambu-Jona-Lasinio models. So far, there are various phenomenological proposals suggesting universal conditions describing the chemical freezeout boundary. For a recent review, the readers are kindly advised to consult Ref. \cite{r4}. Recently, the possible interrelations among the various freezeout conditions have been derived \cite{Tawfik:2016jzk}.

While the region of the high temperature and low baryonic density at the QCD phase-diagram are explored by the experiments of the RHIC and the LHC, the region of relatively low and intermediate energy will be covered by the future programs: BES-II at RHIC, Nuclotron-based Ion Collider fAcility (NICA) at the Joint Institute for Nuclear Research (Dubna) and the Facility for Antiproton and Ion Research (FAIR, Germany). At NICA will be the fixed target experiment BM@N with the beam energy $E_{kin} = 1 - 4.5$ AGeV and the collider experiment MPD  with the collision energy range $4\, \leq \sqrt{s_{NN}} \leq 11\, $ GeV. At the FAIR will work the fixed target experiment CBM with  $E_{kin}$ up to $11$ AGeV (SIS100).

In the present work we utilize  the hadron resonance gas (HRG) \cite{Tawfik:2016jzk} and the Ultra-relativistic Quantum Molecular Dynamics (UrQMD)  v.$3.4$ models \cite{urqmd-34} in order to estimate various particle ratios at energies ranging  for $\sqrt{s_{NN}}$  from $3$ to $19.6~$ GeV. The freezeout parameters ($T_{\mathrm{ch}}$ and $\mu_{\mathrm{b}}$) are determined from the statistical fit of various particle ratios from UrQMD simulations of {Au--Au} collisions at  $\sqrt{s_{NN}} = 3, 5, 7.7, 11.5$ and $19.6$ GeV. The last three energies are the part of the RHIC beam energy scan program (BES I)   and at  these  energies the experimental values of freezout parameters are obtained also. A  comparison of simulated and  these experimental results are discussed. The convincing agreement between UrQMD simulated and experimental parameters  encourages us to extend the study to the other beam energies through UrQMD simulations lower down to $3$ GeV, in which the baryon density likely reaches its maximum and shall be covered by NICA and FAIR future facilities.

It is obvious that HRG is an effective statistical model which is only applicable to the produced particles in their final stages of the temporal and spatial evolution of the high-energy collision. Thus, both chiral and deconfinement phase transition(s) can not be modelled in such statistical approaches, which are based on Hagedorn bootstrap picture \cite{Tawfik:2016jzk}.

The present paper is organized as follows. Section \ref{models} gives short reviews on both approaches; HRG and UrQMD models. In Section \ref{results}, the energy dependence of  various particle-ratios (section \ref{sec:energyratio}) and the deduced freezeout parameters (section \ref{sec:fit}) are presented. In Section \ref{conc}, the conclusions are outlined.

\section{Approaches}
\label{models}

The hybrid UrQMD model is used to  calculate various particle ratios at energies ranging $\sqrt{s_{\mathrm{NN}}}$ from  3 to 19.6~GeV. This is appropriate as long as the location of the critical endpoint is not known yet. It widely varies in both $\mu$- and $T$-dimension. The freezeout parameters: temperature ($T$) and baryon chemical potential ($\mu_{\mathrm{b}}$)  are calculated from the HRG model and the statistical fit of the particle ratios  from  the UrQMD data. This data was generated  with the  first order or crossover phase transitions. Our UrQMD ensemble contains $10\, 000$ and $150\, 000$ events for high and low energies, respectively.  


\subsection{Hadron Resonance Gas (HRG) model}
\label{sec:hrg}

In grand-canonical ensemble, the partition function of an ideal gas consisting of hadrons and resonances is given as \cite{r4}
\begin{equation} \label{GrindEQ1}
 Z(T,V,\mu)=\mbox{Tr}\left[\exp\left(\frac{{\mu}N-H}{T}\right)\right],
\end{equation}
where $H$ is the Hamiltonian and $\mu$ and $T$ being chemical potential  and temperature of the system of interest, respectively. The Hamiltonian counts the relevant degrees of freedom of  confined and strongly interacting medium. Interactions (correlations) can be included implicitly, for instance, in the ones responsible for the resonance formation, i.e., strong interactions. In the HRG model, Eq. (\ref{GrindEQ1}) sums up contributions from a large number of hadron resonances  \cite{r4} consisting of light and strange quark flavors as listed in the most recent particle data group compilation with masses $\leq 2~$GeV \cite{pdg}. This corresponds to $388$ different states of mesons and baryons besides their anti-particles. More details can be found in Ref. \cite{r4}. The decay branching ratios are also taken into consideration. For the decay channels with not-yet-measured probabilities, we follow the rules given in Ref. \cite{r32}. No finite size correction was applied \cite{r6} 
 \begin{equation} \label{GrindEQ2}
{\ln  Z(T,V,\mu)\ }=\sum_i{{\ln Z}_i(T,V,\mu)} = \pm \frac{V g_i}{2 {\pi }^2} \int^{\infty}_0  p^2 dp\;  \ln \left[1\pm \lambda_i \exp\left(\frac{-{\varepsilon}_i(p)}{T} \right) \right],
\end{equation}
where $\pm $ represent  fermions and bosons, respectively, $\varepsilon _{i} =\sqrt{p^{2} +m_{i}^{2} }$ is the dispersion relation of the $i$-th particle and $\lambda_i$ is its fugacity factor \cite{r4}
\begin{equation} \label{GrindEQ4}
\lambda_{i} (T,\mu)=\exp\left(\frac{b_{i} \mu_{\mathrm{b}} +S_{i} \mu_{\mathrm{S}} +Q_{i} \mu_{\mathrm{Q}}}{T} \right).
\end{equation}
where  $b_{i} (\mu _{\mathrm{b}})$, $S_{i} (\mu _{\mathrm{S}})$ and $Q_{i} (\mu _{\mathrm{Q}})$ are baryon, strange and charge quantum numbers (their corresponding chemical potentials) of the $i$-th hadron, respectively.

The number density of $i$-particle can be derived from the derivative with respect to the chemical potential of the corresponding quantum number. Such particle can be a {\it stable} hadron and decay product out of heavy resonances,
\begin{equation} \label{GrindEQ6}
n_i(T,\mu)=
\frac{g_{i}}{2\pi^{2}}\int_{0}^{\infty }\frac{p^{2} dp}{\exp\left[\frac{{\mu}_{i} - {\varepsilon}_{i}(p)}{T}\right] \pm 1} 
+ \sum_{j\neq i} b_{j\rightarrow i} \frac{g_{j}}{2\pi^{2}}\int_{0}^{\infty }\frac{p^{2} dp}{\exp\left[\frac{{\mu}_{j} - {\varepsilon}_{j}(p)}{T}\right] \pm 1},
\end{equation}
where $b_{j\rightarrow i}$ is the decay branching ratio of $j$-th hadron resonance into $i$-th stable particle of antiparticle.
In a statistical fit of various particle ratios with the UrQMD simulations or with the measurements at different energies, $T$ and $\mu_{\mathrm{b}}$ are taken as free parameters. Details about the statistical fit can be taken from Refs. \cite{Tawfik:2013bza,Tawfik:2014dha}. Their resulting values can be related to each other and each of them to the center-of-mass energy ($\sqrt{s_{\mathrm{NN}}}$), separately \cite{r4}.

It noteworthy emphasizing where or when HRG can be applied. As mentioned, different numerical methods and algorithms seems to fail while trying to reproduce even the well-identified particles (low-lying states, such as pions, Kaons and protons) at very low beam energies or equivalently very large baryon chemical potentials. In this energy limit, some particle species can't be accessed. This becomes illustrated in section \ref{results}, especially Fig. \ref{fig:firstorder}. In general, the HRG model is a very powerful statistical approach. In spite of its simplicity, it finds so-far a wide  range of implications, especially in describing various aspects of the lattice QCD thermodynamics and the particle production in heavy-ion collisions. The latter are limited to the final state, post the chemical freezeout era. All prior eras of the relativistic collision are not accessible by the HRG model. These would be subject to transport approaches. The UrQMD characterizes almost the entire evolution of the colliding system from very early stages up to the particle production,  including hydrodynamical evolution and particlization. This {\it transport} approach will be elaborated in the section that follows.

\subsection{Ultrarelativistic Quantum Molecular Dynamics (UrQMD) model}
\label{sec:urqmd}

The UrQMD event-generator \cite{r26} is a well-known simulation approach enabling the characterization of high-energy collisions. It simulates the phase space of such collisions and implements a large set of Monte Carlo solutions for a large number of paired partial-differential equations describing the evolution of phase space densities. The UrQMD model simulates the development of the colliding system from a possibly very early stage (depending on the chosen configuration) up to the final state of the particle production. Its large number of unknown parameters could be fixed from experimental results and by theoretical assumptions. 

In the present calculations, we use hybrid UrQMD~v.$\,3.4$~\cite{urqmd-34} which  has been tested and give reasonable results in the energy range from $E_{lab} = 2 - 160~$AGeV in standard parameter calculations. Furthermore,  hybrid UrQMD v$3.4$ provides the possibility to use two types of the phase transition; first order and crossover. This allows us the study of the possible effects of the hadronization processes on the final-state particle production.


{\color{red}
In hybrid UrQMD $3.4$,  for case the of crossover, the equation of state (EoS) for the fluid dynamical evolution is borrowed from the SU($3$) parity doublet model in which quark degrees-of-freedom besides the thermal contribution of the Polyakov loop are included \cite{sss,sdbp}. This EoS  qualitatively agrees with the lattice QCD results at vanishing baryon chemical potential and - most importantly -  is conjectured to be applicable at finite baryon chemical potentials, as well.  For first-order phase transition, an EoS from SU($2$) bag model is to be included. By the end of the hydrodynamical evolution, the active EoS is changed to the one characterizing the hadron gas. Accordingly, it is assured that the active degrees-of-freedom on both sides of the transition hypersurface are exactly equivalent \cite{sss,sdbp,AuvPet}. 
}

For the first order phase transition in UrQMD~v.$\, 3.4$ is used the approach proposed in~\cite{r28}. The nuclear matter is described by a $ \sigma - \omega~$-type model for the hadron matter phase and by the MIT bag model for the quark-gluon plasma, with a first order phase transition between both phases.

For the sake of completeness, we emphasize that two differences between crossover and first-order phase transitions are the latent heat and the degrees of freedom.  In first-order phase transition both are larger than that in crossover. Furthermore, the crossover takes place smoothly, i.e., a relative wide range of temperatures is needed to convert the QCD matter from pure hadron to parton matter or vice versa, while there is a prompt jump in case of first-order phase transition, i.e., the critical temperature becomes very sharp \cite{r31}.

As a limitation, there is some influence of a mere technical aspect of UrQMD to its physical outcome. When the program switches from the hydrodynamical treatment of the high-density stage of the hadronic medium back to the {\it ``normal''} particle-based transport code, there might occur some bias to the resulting particle statistics \cite{particlization}. Because we selected the same particlization procedure in both cases, the differences of particle ratios between first order and crossover phase might appear smaller than implied by the physical model.

\section{Results and Discussion}
\label{results}

 Particles ratios in this work are studied  at the energies $\sqrt{s_{\mathrm{NN}}} = 3, 5, 7, 7.7, 9, 11, 11.5, 13, 19$ and $19.6~$GeV. The energies $7.7$, $11.5$ and $19.6~$GeV are the part of the STAR BES program and for these energies  exist also comparable measurements from experiments of the Superproton synchrotron (SPS), such as NA49, NA44, and NA57 \cite{r32}. 
The energy range  $\sqrt{s_{\mathrm{NN}}} = 3 - 11~$GeV will be reached at future NICA facility, so it is interesting to study particle ratios  in both regions experimentally and with the model approaches. If the hybrid UrQMD model can reproduce the STAR results relatively well further UrQMD simulations for the energies  $3$ and $5~$ GeV will be as predictions for the future experiments at NICA.

From an ensemble of  events created by the hybrid UrQMD model at various energies and taking into account two types of the quark-hadron phase transition (crossover and first order), we study the ratios of various particle species. For beginning, we determine their energy dependence.  From the statistical fit of the HRG model to the ones simulated with UrQMD and independently to the data from STAR experiments, both freezeout parameters were calculated.


\subsection{Energy dependence of various particle ratios}
\label{sec:energyratio}

\begin{figure}[htb]
\includegraphics[width=8cm]{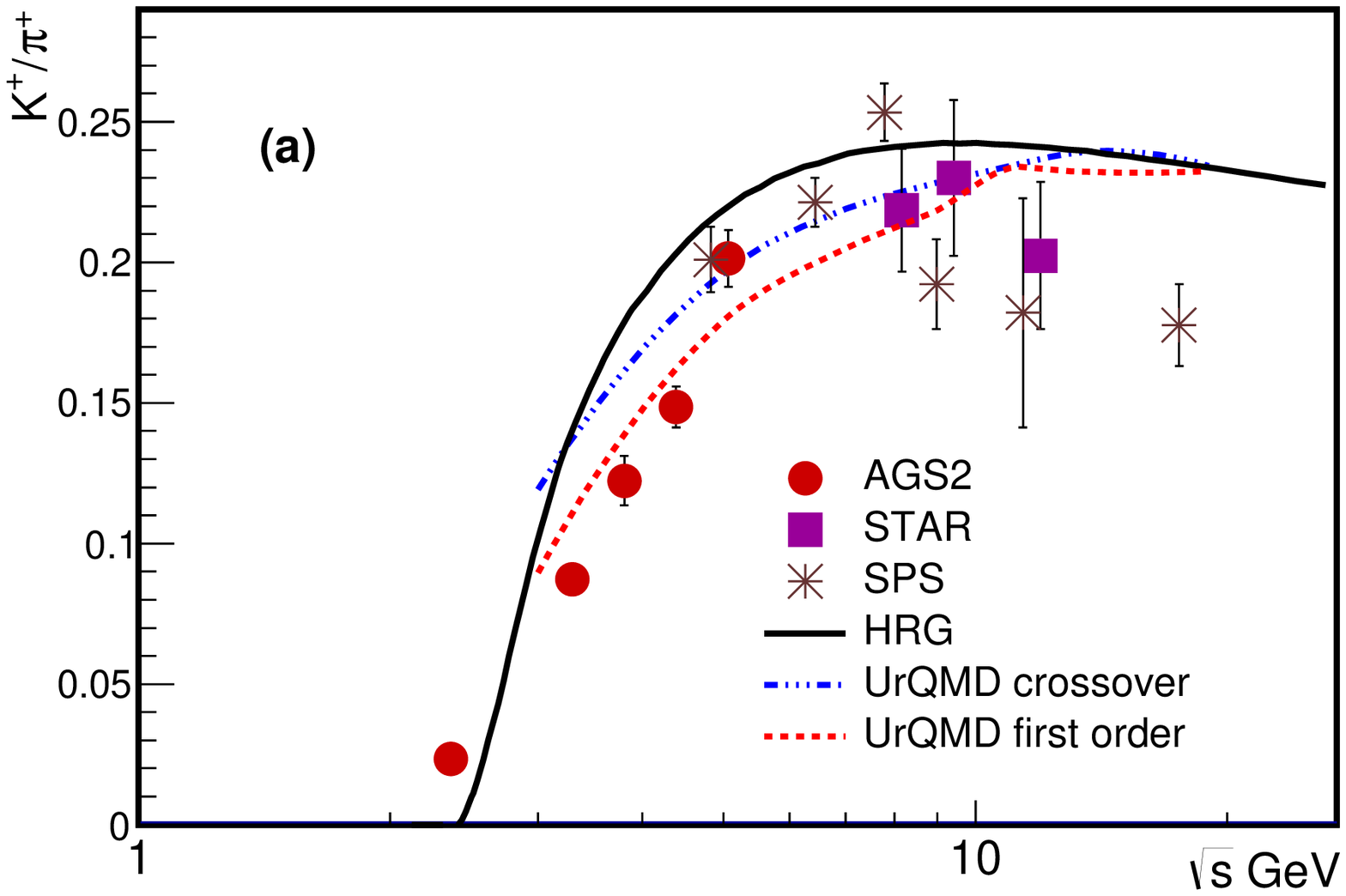}
\includegraphics[width=8cm]{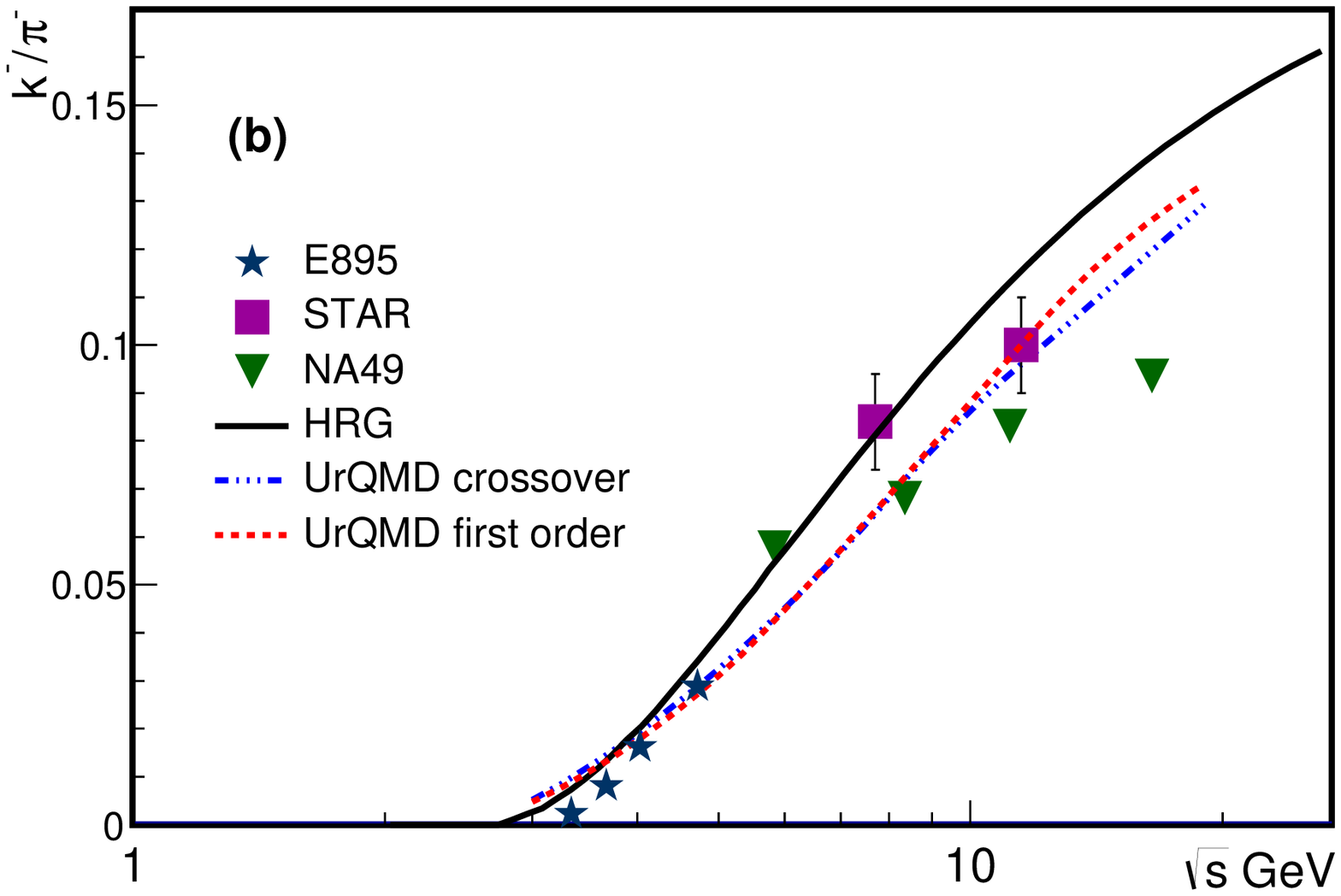}
\includegraphics[width=8cm]{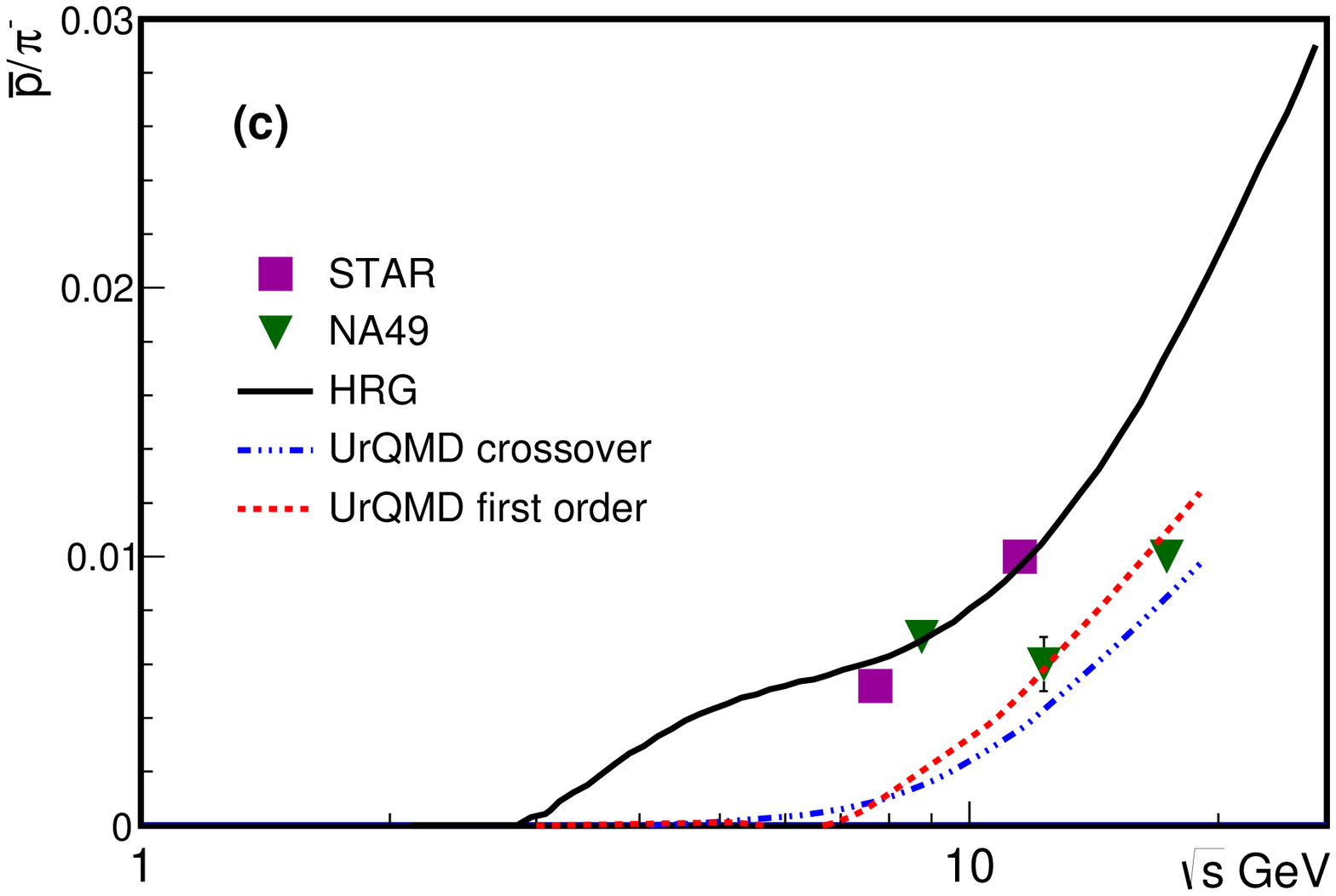}
\includegraphics[width=8cm]{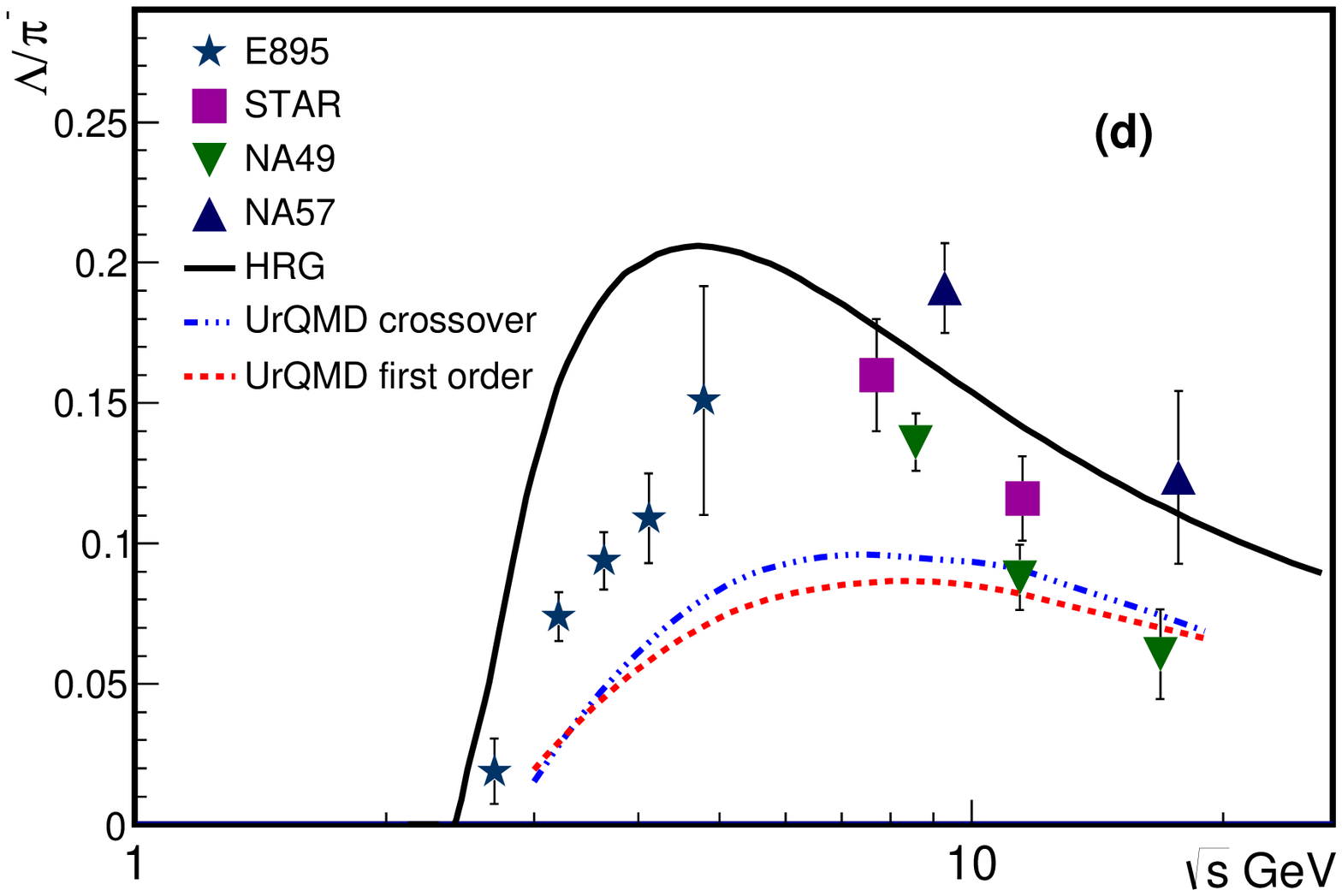}
\includegraphics[width=8cm]{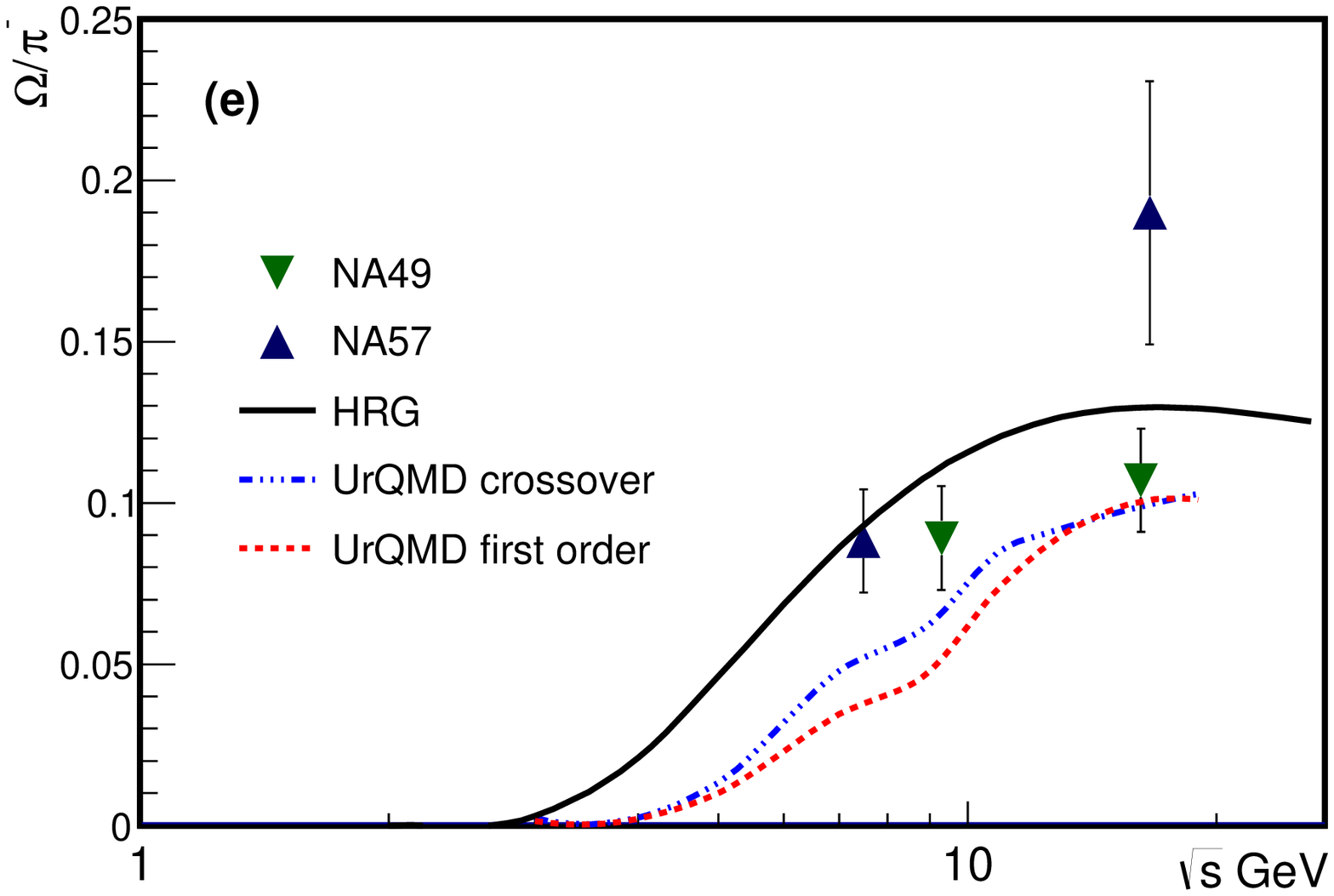}
\caption{The energy dependence of different particle ratios calculated from hybrid UrQMD with first-order (dashed curves) and crossover (triple-dot-dashed curves) phase transition(s) is compared with various measurements:  $\mathrm{K}^+/\mathrm{\pi}^+$ (a) \cite {r47,r41,r49}, $\mathrm{K}^-/\mathrm{\pi}^-$  (b) \cite {r47,r41,r49,r38}, $\mathrm{\bar{p}}/\mathrm{\pi}^-$  (c) \cite {r47,r41,r49,r43}, $\mathrm{\Lambda}/\mathrm{\pi}^-$   (d) \cite{r47,r41,r49,r38,r33,r50,r40},   $\mathrm{\Omega}/\mathrm{\pi}^-$ (e) \cite {r47,r41,r49,r38,r35,r40,r50}. The solid curve represent the corresponding calculations from the HRG model.  }
\label{fig:one}
\end{figure}

\begin{figure}[htb]
\includegraphics[width=8cm]{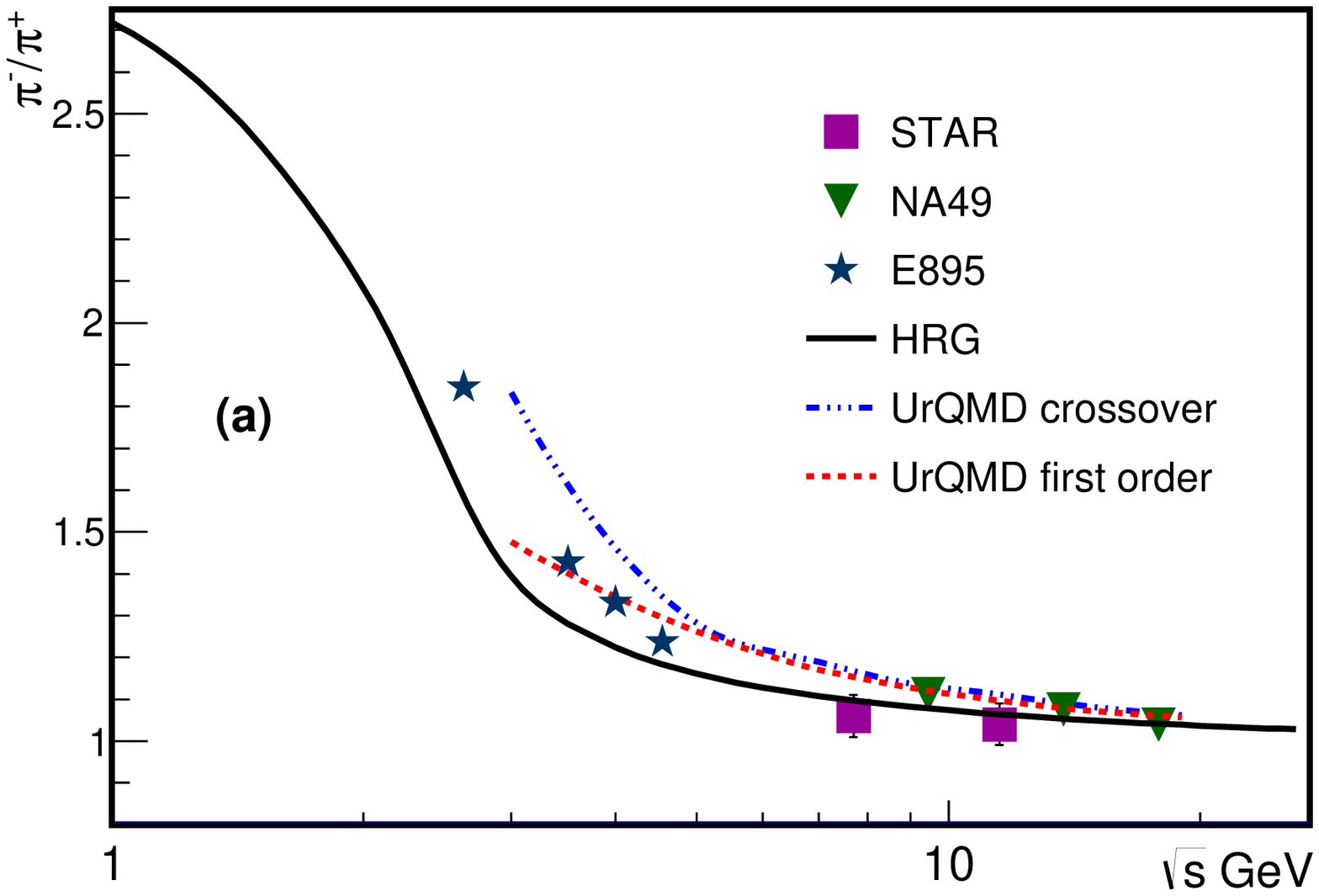}
\includegraphics[width=8cm]{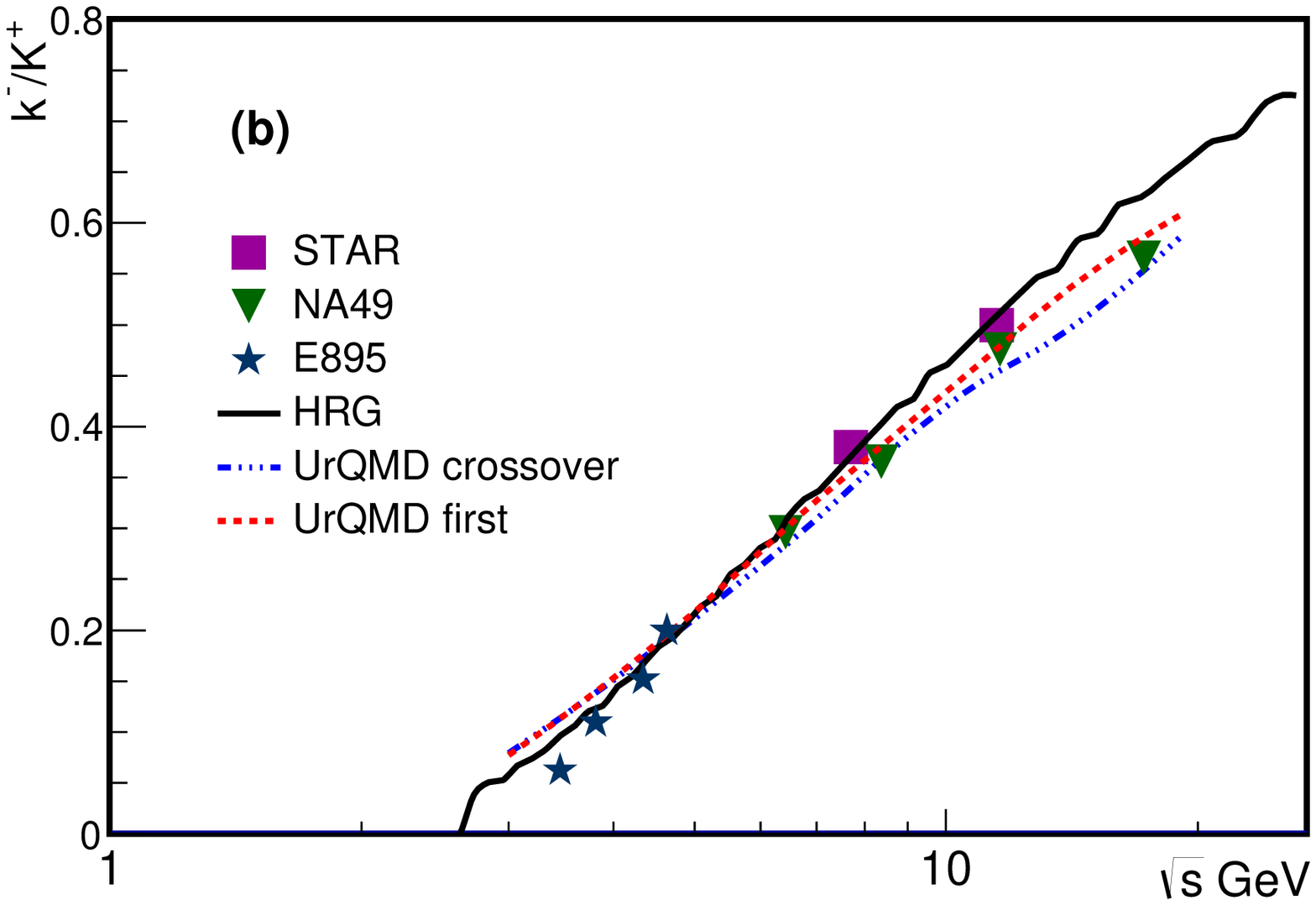}
\includegraphics[width=8cm]{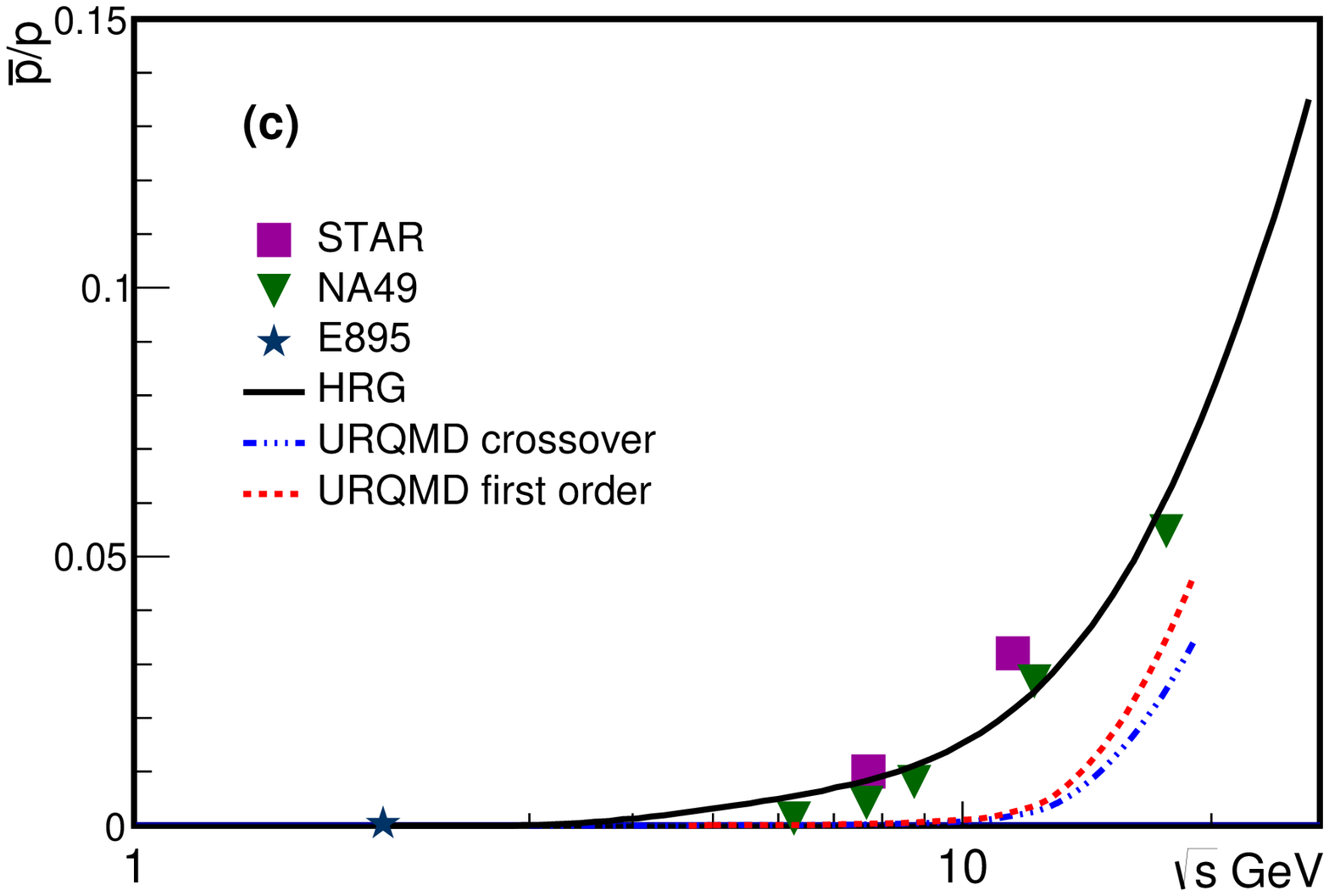}
\includegraphics[width=8cm]{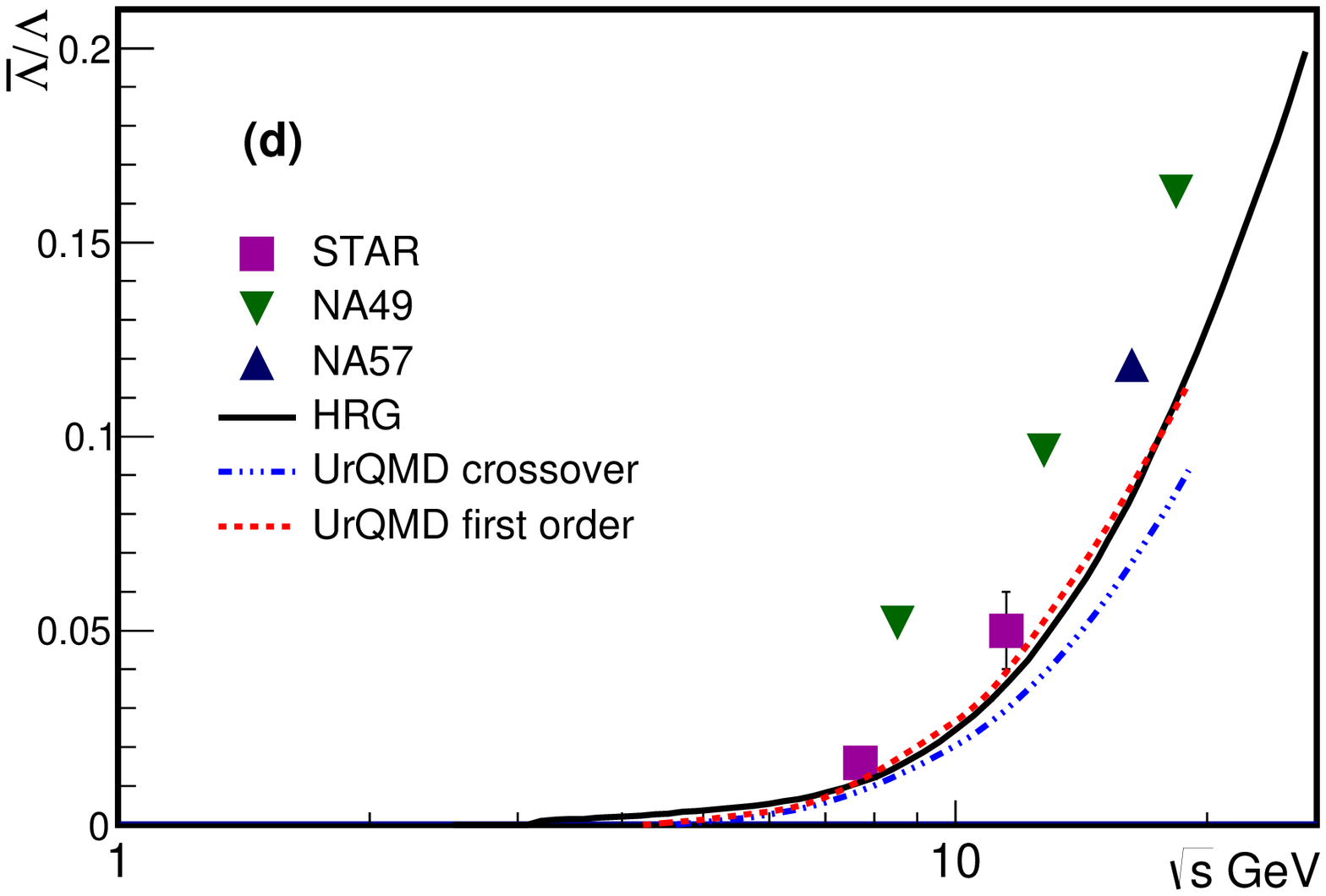}
\includegraphics[width=8cm]{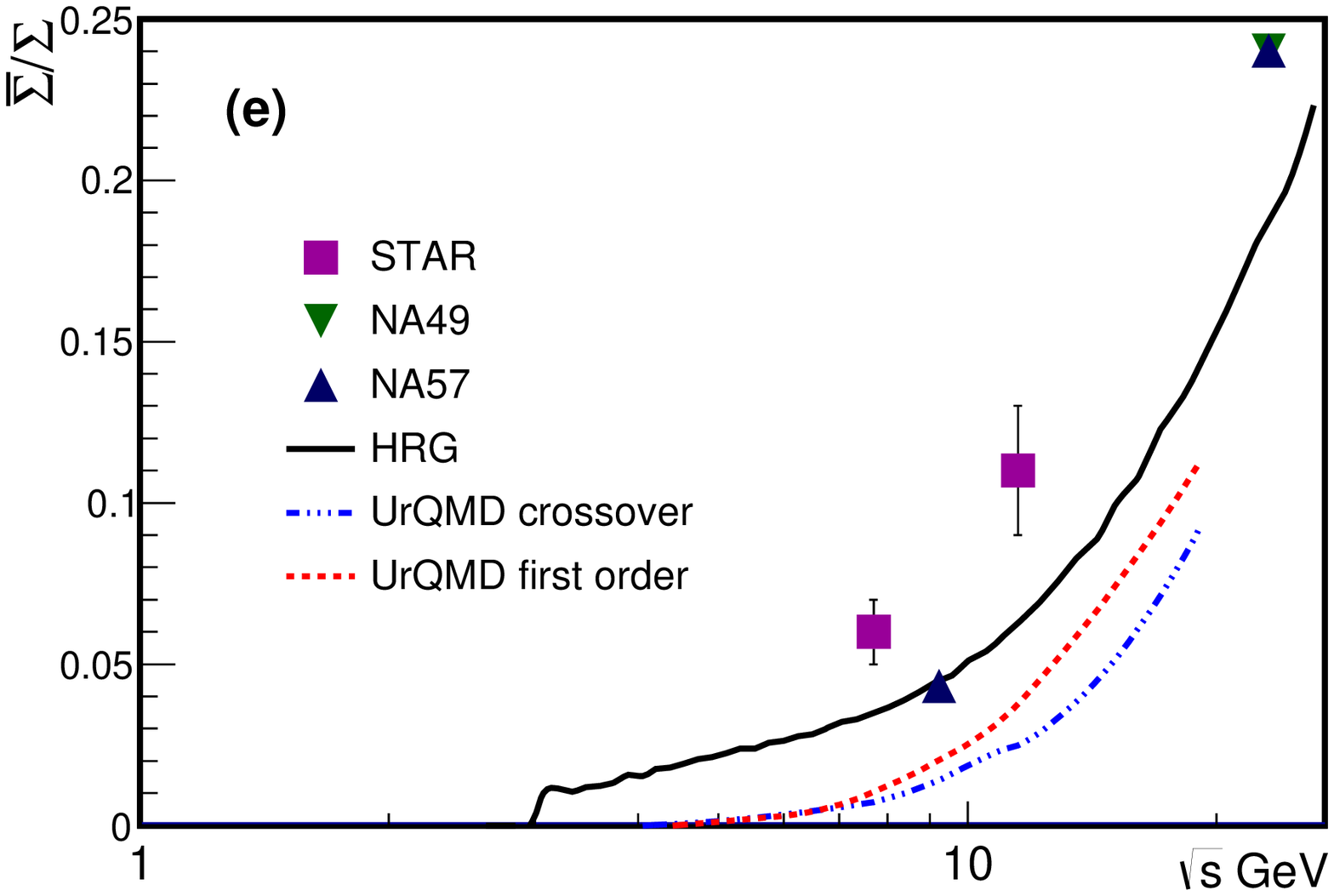}
\caption{The same as in Fig. \ref{fig:one} but for $\mathrm{\pi}^-/\mathrm{\pi}^+$ (a)  \cite{r47,r41,r49,r38}, $\mathrm{K}^-/\mathrm{K}^+$ (b) \cite{r47,r41,r49,r38}, $\mathrm{\bar{p}}/\mathrm{p}$ (c) \cite{r47,r41,r49,r43}, $\bar{\Lambda}/\Lambda$ (d) \cite{r47,r41,r49,r33,r50,r40,r38} and  $\bar{\Sigma}/\Sigma$ (e) \cite{r34,r36,r52}. }
\label{fig:two}
\end{figure}

Fig. \ref{fig:one} presents the energy dependence of the particle ratios $\mathrm{K}^+/\mathrm{\pi}^+$ (a), $\mathrm{K}^-/\mathrm{\pi}^-$ (b), $\mathrm{\bar{p}}/\mathrm{\pi}^-$ (c), $\mathrm{\Lambda}/\mathrm{\pi}^-$ (d) and $\mathrm{\Omega}/\mathrm{\pi}^-$ (e). They are generated from the hybrid UrQMD event-generator at various energies and by taking into account two types of the quark-hadron phase transition [crossover (triple-dot-dashed curves) and first order (dashed curves)]. The UrQMD results are compared with measurements (symbols): $\mathrm{K}^+/\mathrm{\pi}^+$ (a) \cite{r47,r41,r49}, $\mathrm{K}^-/\mathrm{\pi}^-$  (b) \cite {r47,r41,r49,r38}, $\mathrm{\bar{p}}/\mathrm{\pi}^-$  (c) \cite {r47,r41,r49,r43}, $\mathrm{\Lambda}/\mathrm{\pi}^-$   (d) \cite{r47,r41,r49,r38,r33,r50,r40},   $\mathrm{\Omega}/\mathrm{\pi}^-$ (e) \cite {r47,r41,r49,r38,r50,r40,r35} and the corresponding calculations from the HRG model.

The HRG particle ratios are determined from Eq. (\ref{GrindEQ6}), in which the baryon chemical potential ($\mu_{\mathrm{b}}$) is replaced by $\sqrt{s_{\mathrm{NN}}}$ \cite{Tawfik:2013bza}
\bea
\mu_{\mathrm{b}} &=& \frac{a}{1+b\, \sqrt{s_{\mathrm{NN}}}}, \label{eq:musqrtsNN}
\eea
where $a=1.245\pm 0.094~$GeV and $b=0.264\pm 0.028~$GeV$^{-1}$.
The HRG calculations are in good agreement with both measurements and UrQMD predictions, at least qualitatively. For some of the particle ratios, the agreement is better than for the other ratios. It should be noticed, that these calculations will be fine-tuned in order to reproduce both UrQMD and the experimental results. In doing this, both freezeout parameters will be taken as free variables. Adjusting both of the parameters brings HRG calculations to a quantitative agreement with the UrQMD and the experimental results. It is worthwhile noticing that the particle ratios from both types of phase transition are almost indistinguishable, especially at lower energies (larger baryon chemical potentials).

In Fig. \ref{fig:two}, the energy dependence of UrQMD $\mathrm{\pi}^-/\mathrm{\pi}^+$, $\mathrm{K}^-/\mathrm{K}^+$, $\mathrm{\bar{p}}/\mathrm{p}$, $\mathrm{\bar{\Lambda}}/\mathrm{\Lambda}$ and  $\mathrm{\bar{\Sigma}}/\mathrm{\Sigma}$ are illustrated and compared with HRG calculations and various  measured ratios; $\mathrm{\pi}^-/\mathrm{\pi}^+$ (a)  \cite{r47,r41,r49,r38}, $\mathrm{K}^-/\mathrm{K}^+$ (b) \cite{r47,r41,r49,r38}, $\mathrm{\bar{p}}/\mathrm{p}$ (c) \cite{r47,r41,r49,r43}, $\bar{\Lambda}/\Lambda$ (d) \cite{r47,r41,r49,r38,r33,r50,r40} and  $\bar{\Sigma}/\Sigma$ (e) \cite{r34,r36,r52}.  Again, it is obvious that the both orders of the phase transitions implemented in the hybrid UrQMD - at least qualitatively - result in both measured (STAR) and calculated (HRG) particle-ratios. Concretely, the particle ratios from crossover phase transition is slightly higher than the ones from the first-order. Furthermore, we observe that the agreement between UrQMD simulations or HRG calculations for these particle-antiparticle ratios and their measurements is fairly convincing, at least qualitatively.

Characterizing the energy dependence of ten particle-ratios from simulations, calculations and measurements, and being successful in reproducing, at least qualitatively, both UrQMD predictions and STAR measurements by means of the statistical-thermal HRG furnish us with a solid argumentation for the attempt to deduce the freezeout parameters from the given data sets. In doing this, we assume the UrQMD simulations take the position of experiments such as STAR. The determination of the freezeout parameters at the energies covered by STAR BES; $7.7$, $11.5$ and $19.6~$GeV are compatible with the UrQMD simulations with crossover phase-transition. The results from the HRG statistical fits well with the STAR BES measurements \cite{r53,r54,r55,r56} can be summarized as
\begin{itemize}
\item at $7.7~$GeV, $T_{\mathrm{ch}}=141~$MeV and $\mu_{\mathrm{b}}=412~$MeV with $\chi^2/{\mathrm{dof}}=11.7/9$,
\item at $11.5~$GeV, $T_{\mathrm{ch}}=150~$MeV and $\mu_{\mathrm{b}}=312~$MeV with $\chi^2/{\mathrm{dof}}=6.9/9$ and
\item at $19.6~$GeV, $T_{\mathrm{ch}}=153~$MeV and $\mu_{\mathrm{b}}=149~$MeV with $\chi^2/{\mathrm{dof}}=7.6/9$.
\end{itemize}
The results from the statistical fit of the HRG calculations to the UrQMD simulations at the considered values of the collision energies can be found in Tab. \ref{tab:1}. It is obvious that both sets of freezeout parameters are compatible with each other.

\subsection{Determining freezeout parameters}
\label{sec:fit}

The study of the energy-dependence of various particle ratios 
paves the way towards determining the freezeout parameters, which are taken as free parameters in the HRG approach, from UrQMD simulations. The statistical fit of the HRG calculations  to the UrQMD results is motivated by the excellent agreement between UrQMD and STAR particle ratios at the given RHIC-BES energies. The quality of the statistical fit is measured by minimum 
$\chi^{2}$ and $q^2$
\begin{equation}
\chi^{2}=\sum_{i}\frac{\left(R_{i}^{\mathrm{exp}}-R_{i}^{\mathrm{theor}}\right)^2}{\sigma^{2}_i}, \qquad\qquad   q^{2}=\sum_{i}\frac{\left(R_{i}^{\mathrm{exp}}-R_{i}^{\mathrm{theor}}\right)^2}{\left(R_{i}^{\mathrm{theor}}\right)^2} \label{GrindEQ12},
\end{equation}
where $R_{i}^{\mathrm{exp}}$ and $R_{i}^{\mathrm{theor}}$ are the $i$-th measured and calculated particle-ratio, respectively, and $\sigma_i$ represents the error in its  measurement. In UrQMD, $\sigma_i$ is restricted to the statistical errors of each particle ratio.

\begin{figure}[htb]
\includegraphics[width=6.5cm,angle=0]{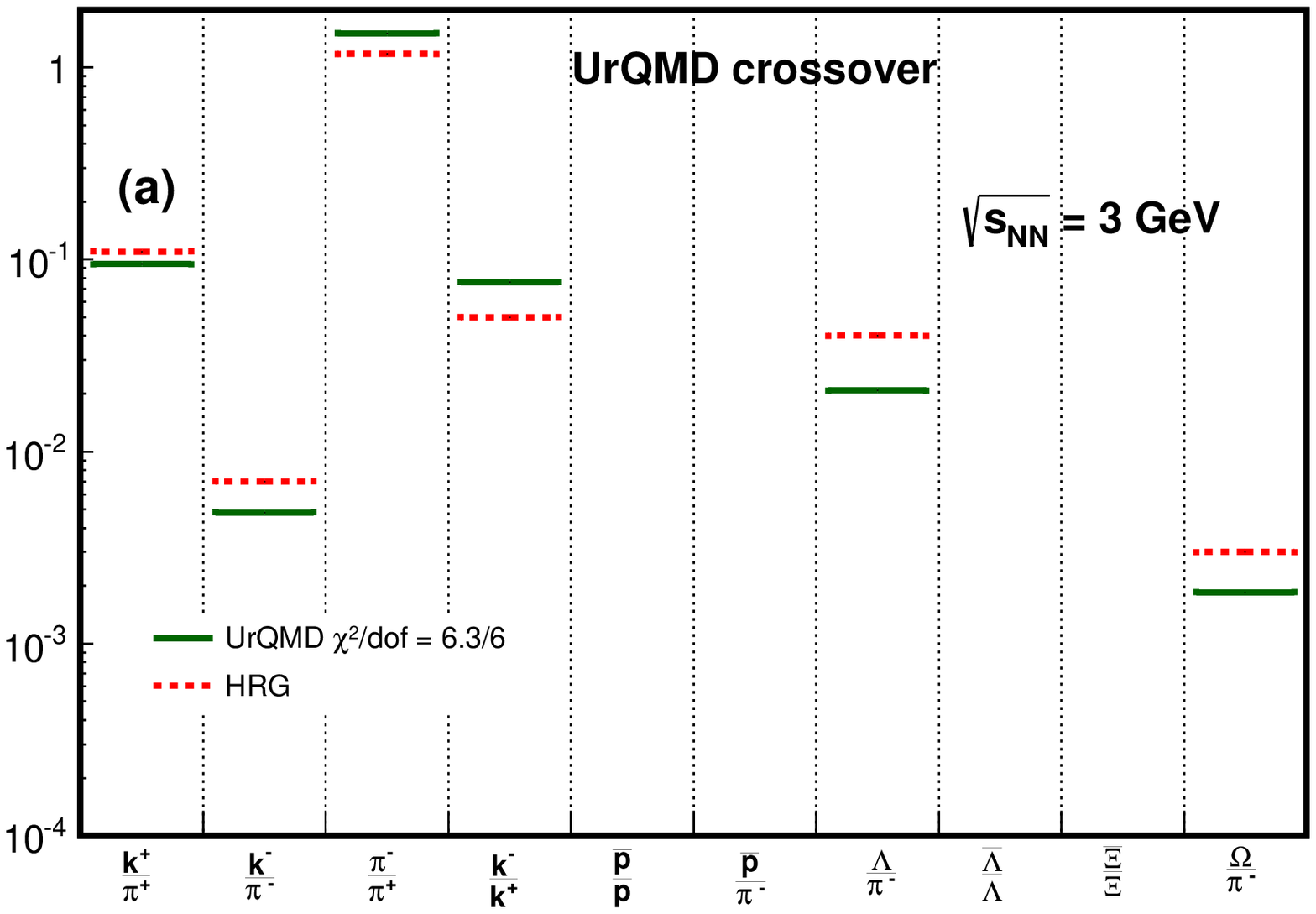}
\includegraphics[width=6.5cm,angle=0]{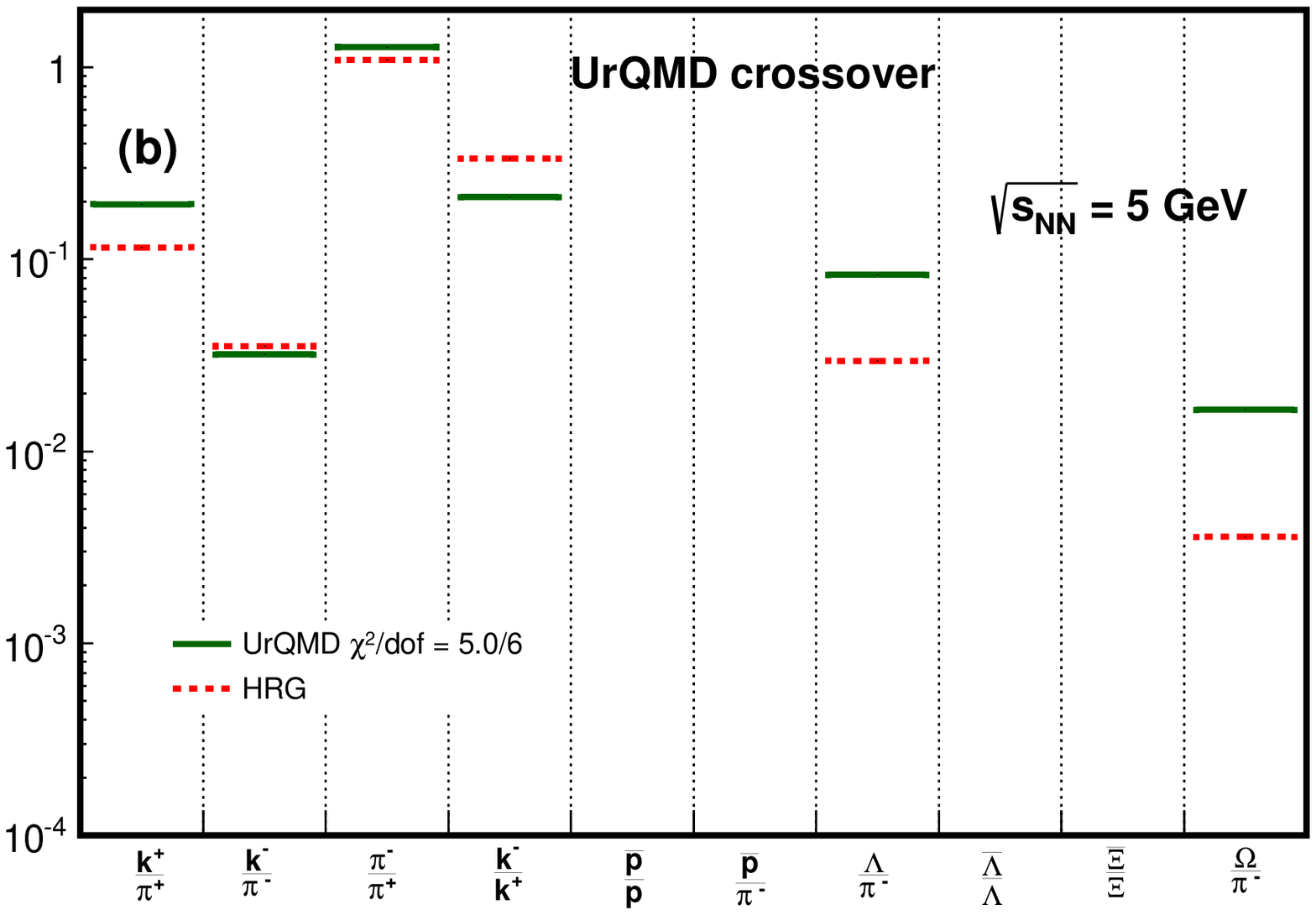}
\includegraphics[width=6.5cm,angle=0]{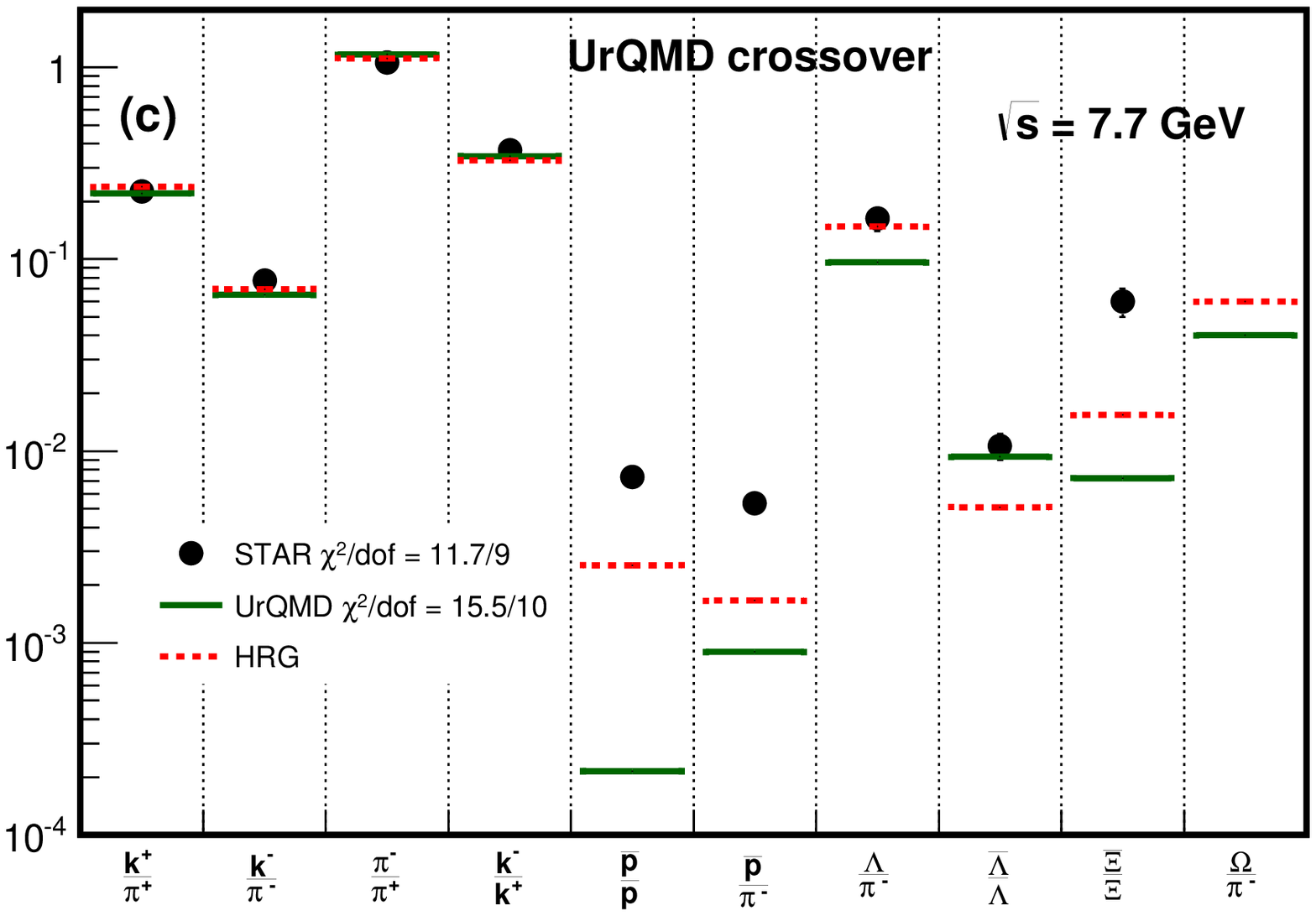}
\includegraphics[width=6.5cm,angle=0]{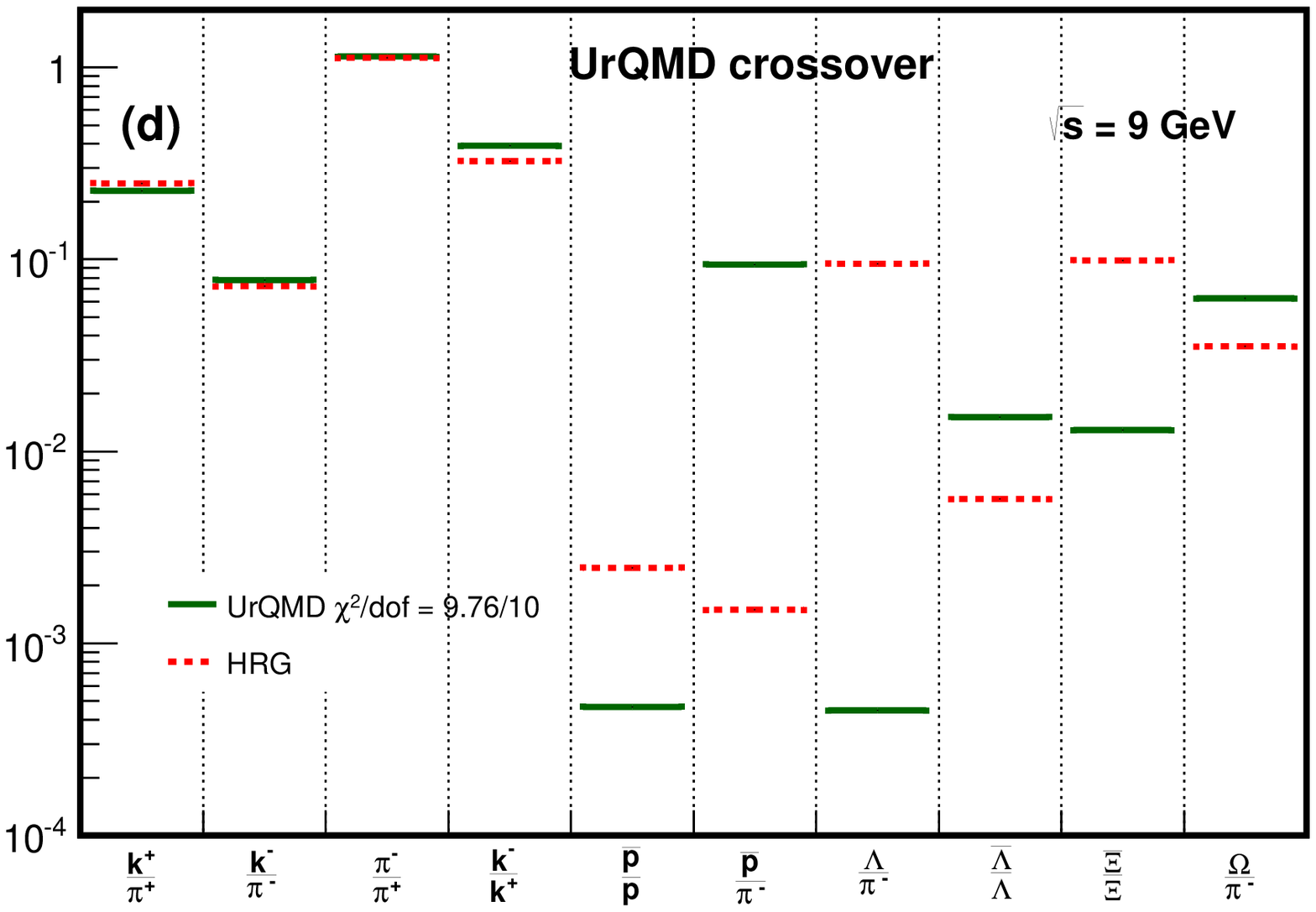}
\includegraphics[width=6.5cm,angle=0]{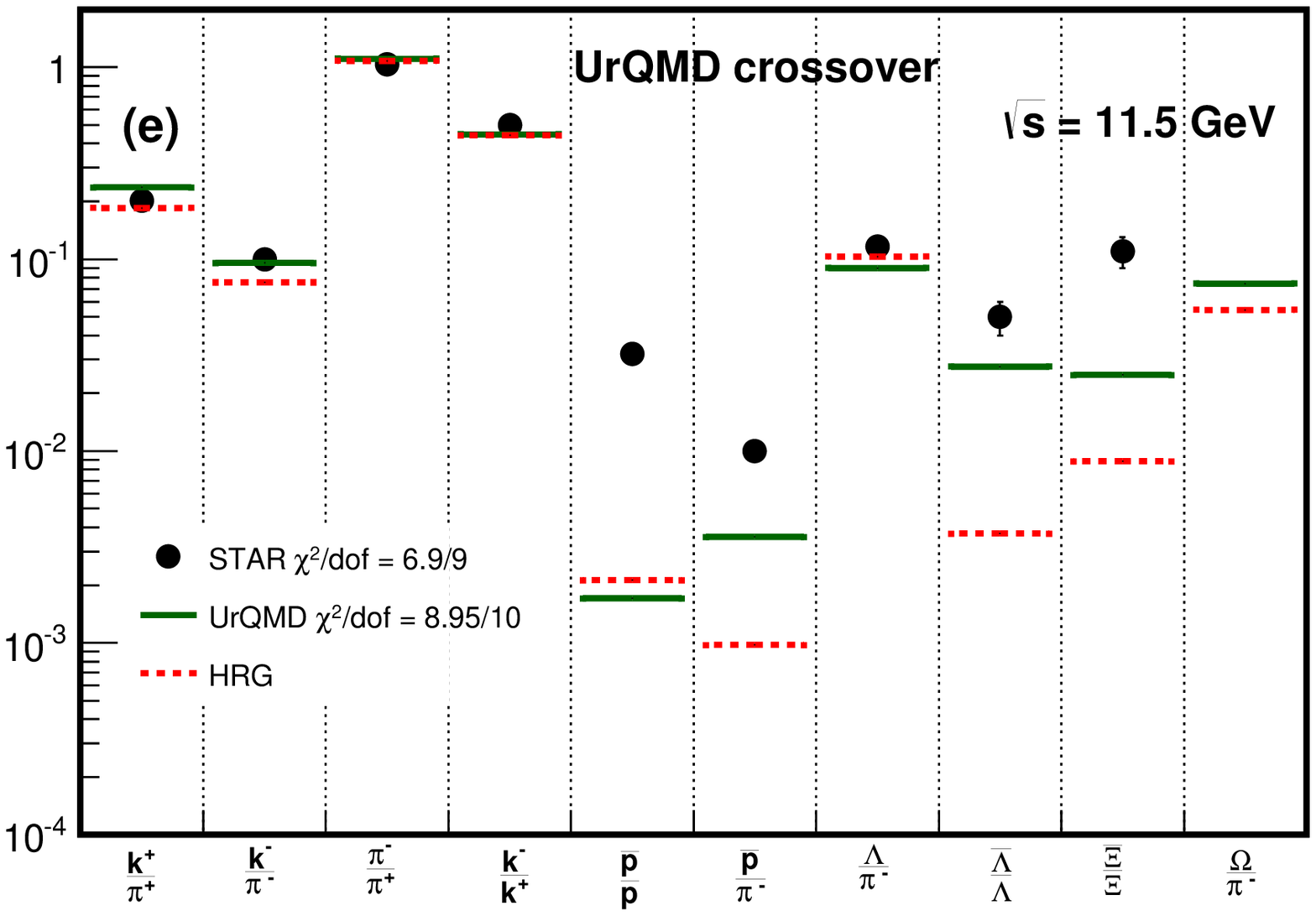}
\includegraphics[width=6.5cm,angle=0]{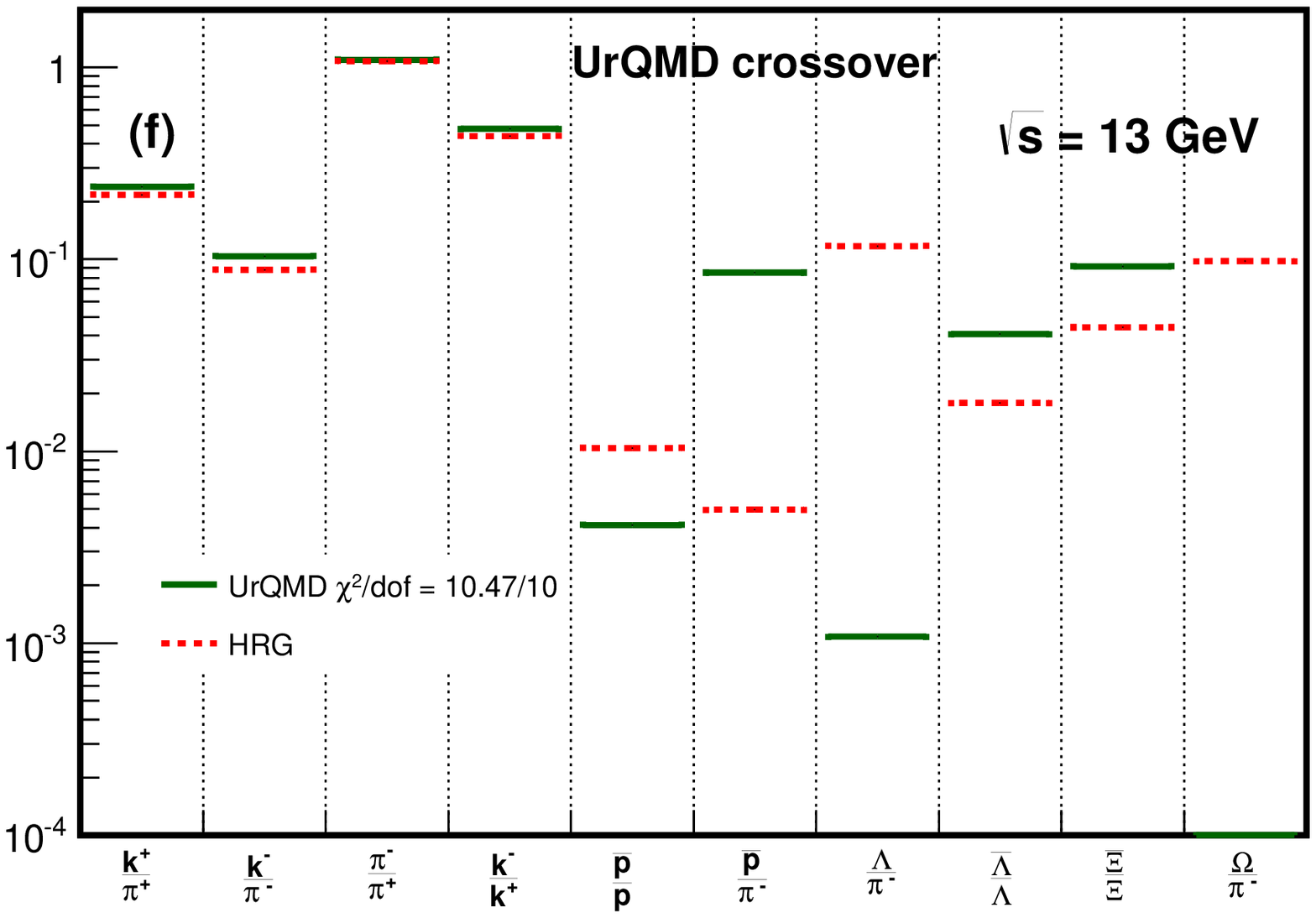}
\includegraphics[width=6.5cm,angle=0]{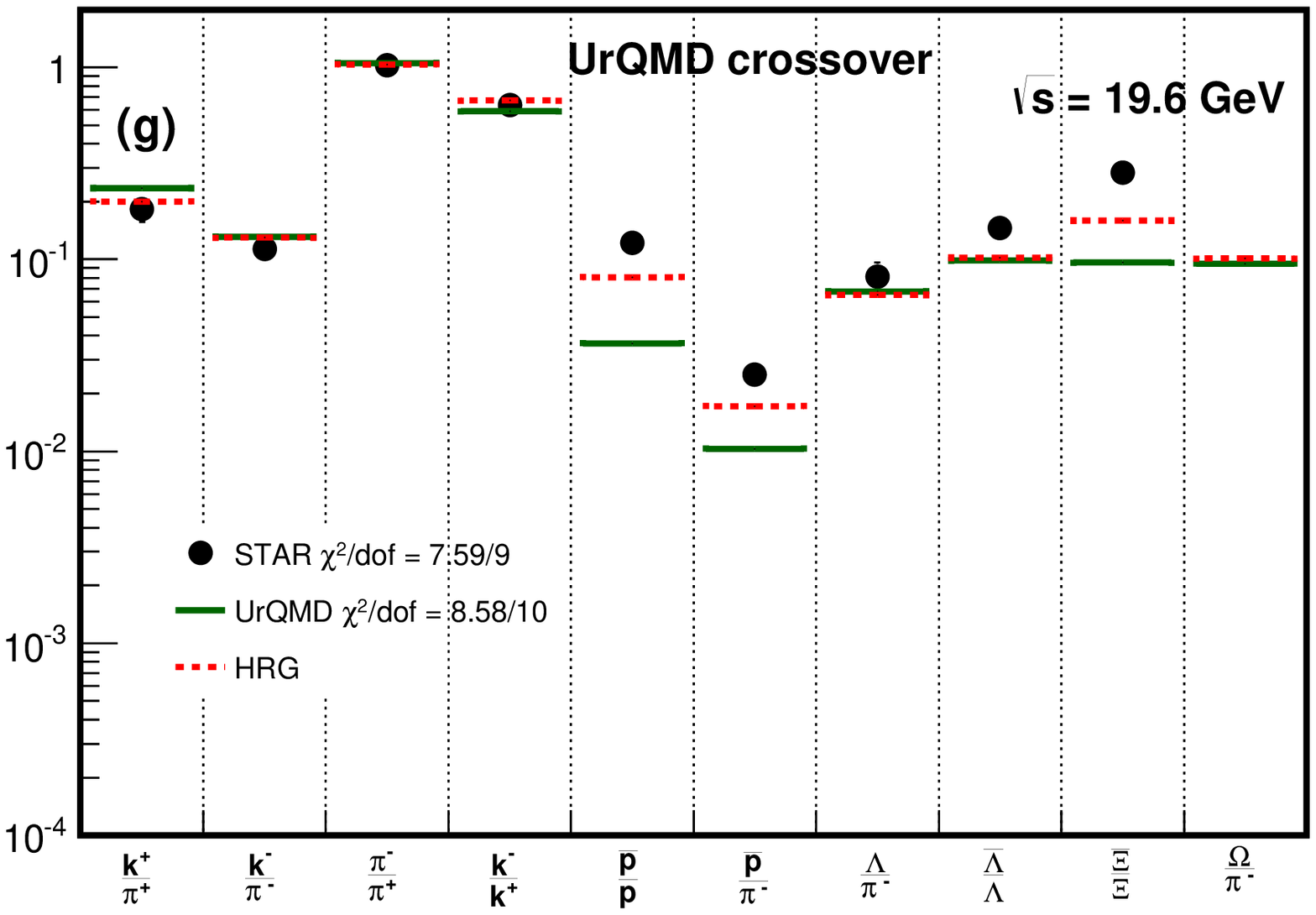}
\caption{Assuming crossover phase transition, UrQMD particle ratios (solid lines) are fitted with the HRG calculations (dashed lines): (a) at $7.7$, (b) at $9$, (c) at $11.5$, (d) at $13$ and (e) at $19.6~$GeV. The resulting freezeout parameters $T_{\mathrm{ch}}$ and $\mu_{\mathrm{b}}$ are listed in Tab. \ref{tab:1}. The open symbols give STAR BES measurements \cite{r53,r54,r55,r56}. The smallest $\chi^{2}/{\mathrm{dof}}$ is given in each graph.
\label{fig:crossover}
}
\end{figure}

For the particle ratios $\mathrm{K}^+/\mathrm{\pi}^+$, $\mathrm{K}^-/\mathrm{\pi}^-$, $\mathrm{\pi}^-/\mathrm{\pi}^+$, $\mathrm{K}^-/\mathrm{K}^+$, $\mathrm{\Lambda}/\mathrm{\pi}^-$, $\mathrm{\bar{p}}/\mathrm{p}$, $\mathrm{\bar{\Lambda}}/\mathrm{\Lambda}$, $\mathrm{\bar{\Sigma}}/\mathrm{\Sigma}$, $\mathrm{\bar{p}}/\mathrm{\pi}^-$ and $\mathrm{\Omega}/\mathrm{\pi}^-$ a comparison between the HRG statistical fits (dashed lines) with the UrQMD simulations with a crossover phase transition (solid lines) and that with the STAR measurements \cite{r53,r54,r55,r56} 
at $7.7$, $11.5$ and $19.6~$GeV (open symbols) is illustrated in panels (a), (c) and (e) of Fig. \ref{fig:crossover}. It is apparent that the ability of hybrid UrQMD to generate the STAR particle ratios increases with the beam energy. This is also reflected in the corresponding  $\chi^{2}$ per degrees of freedom (dof), Tab. \ref{tab:1}. Same observation can be reported from qualities of the HRG statistical fits for both UrQMD and STAR results. The comparison between the hybrid UrQMD simulations and the STAR measurements is illustrated in these three panels in order to argue for further UrQMD simulations at other energies, such as $3$, $5$, $7$, $9$, $11$, $13$ and $19~$GeV.

\begin{table}[htb]
\begin{tabular}{|c|c|c|c|}
\hline
$\sqrt{s_{\mathrm{NN}}}$ [GeV] & $T_{\mathrm{ch}}$ [MeV] & $\mu_{\mathrm{b}}$ [MeV] & $\chi^{2}/{\mathrm{dof}}$ \\
\hline
 3 & 101 & 636 & 6.3/6  \\
\hline
 5 & 109.5 & 599 & 5.0/6  \\
 \hline
  7.7 & 132 & 436 & 15.5/10   \\
\hline
 9 & 132.4 & 429 & 9.76/10 \\
\hline
 11.5 & 138.1 & 391 & 8.95/10  \\
\hline
 13 & 140 & 355 & 10.47/10  \\
\hline
 19.6 & 145 & 200 & 8.58/10  \\
 \hline
\end{tabular}
\caption{Estimated freezeout parameters, $T_{\mathrm{ch}}$ and $\mu_{\mathrm{b}}$ in MeV from the statistical fits of the HRG calculations with the hybrid UrQMD simulations, in which a crossover phase transition is taken into consideration. \label{tab:1} }
\end {table}

\begin{table}[htb]
\begin{tabular}{|c|c|c|c|}
\hline
$\sqrt{s_{\mathrm{NN}}}$ [GeV] & $T_{\mathrm{ch}}$ [MeV] & $\mu_{\mathrm{b}}$ [MeV] & $\chi^{2}/{\mathrm{dof}}$ \\
\hline
 3 & 99 & 630 & 6.4/6  \\
\hline
 5 & 103 & 595 & 5.2/6  \\
\hline
 7 & 125 & 551 & 16.43/10  \\
\hline
 11 & 133 & 395 & 9.87/10  \\
\hline
 19 & 142 & 250 & 8.43/10  \\
\hline
\end{tabular}
\caption{The same as in Tab. \ref{tab:1} but for  hybrid UrQMD simulations with a first-order phase transition. \label{tab:2}}
\end {table}

\begin{figure}[htb]
\includegraphics[width=6.5cm,angle=0]{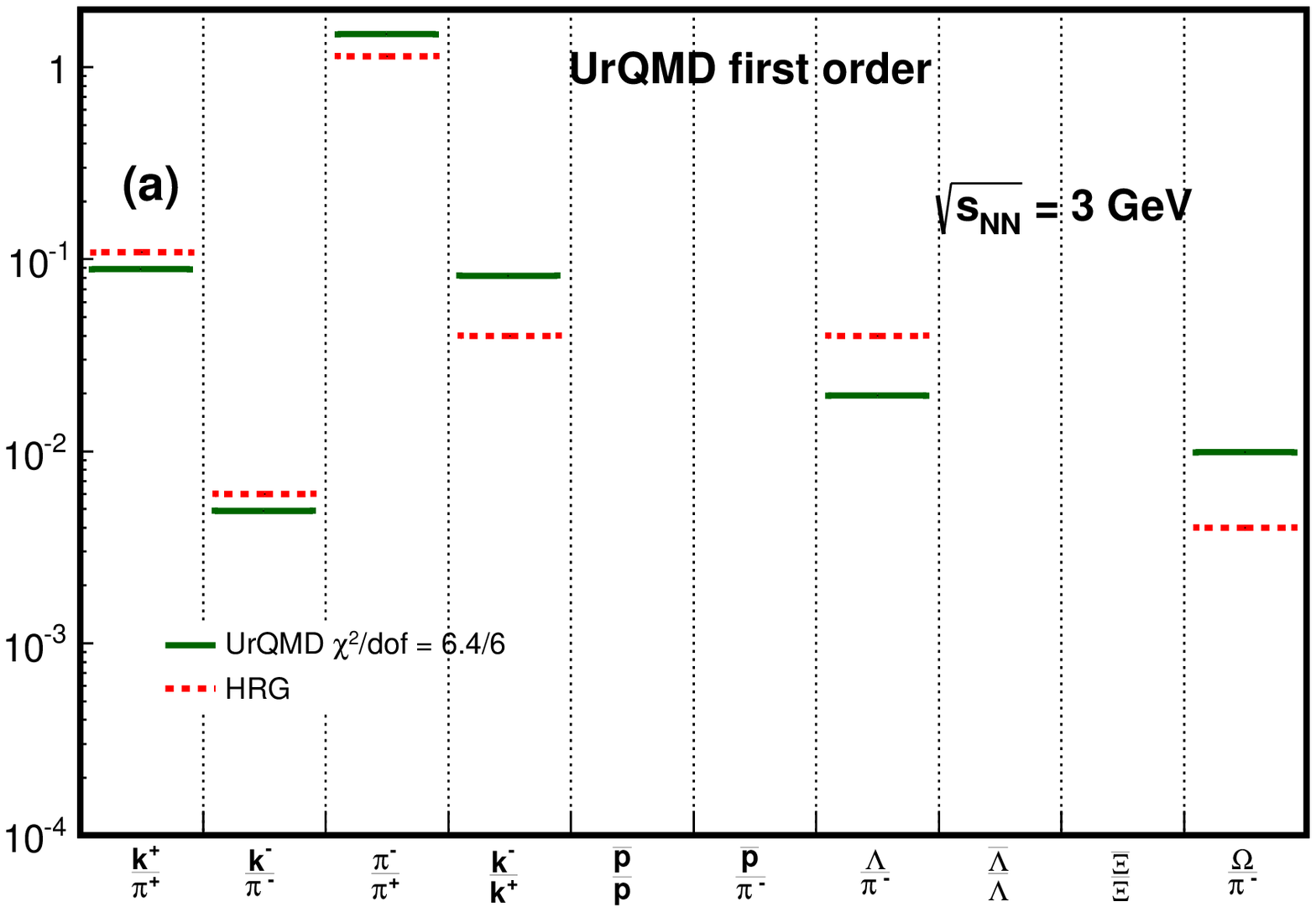}
\includegraphics[width=6.5cm,angle=0]{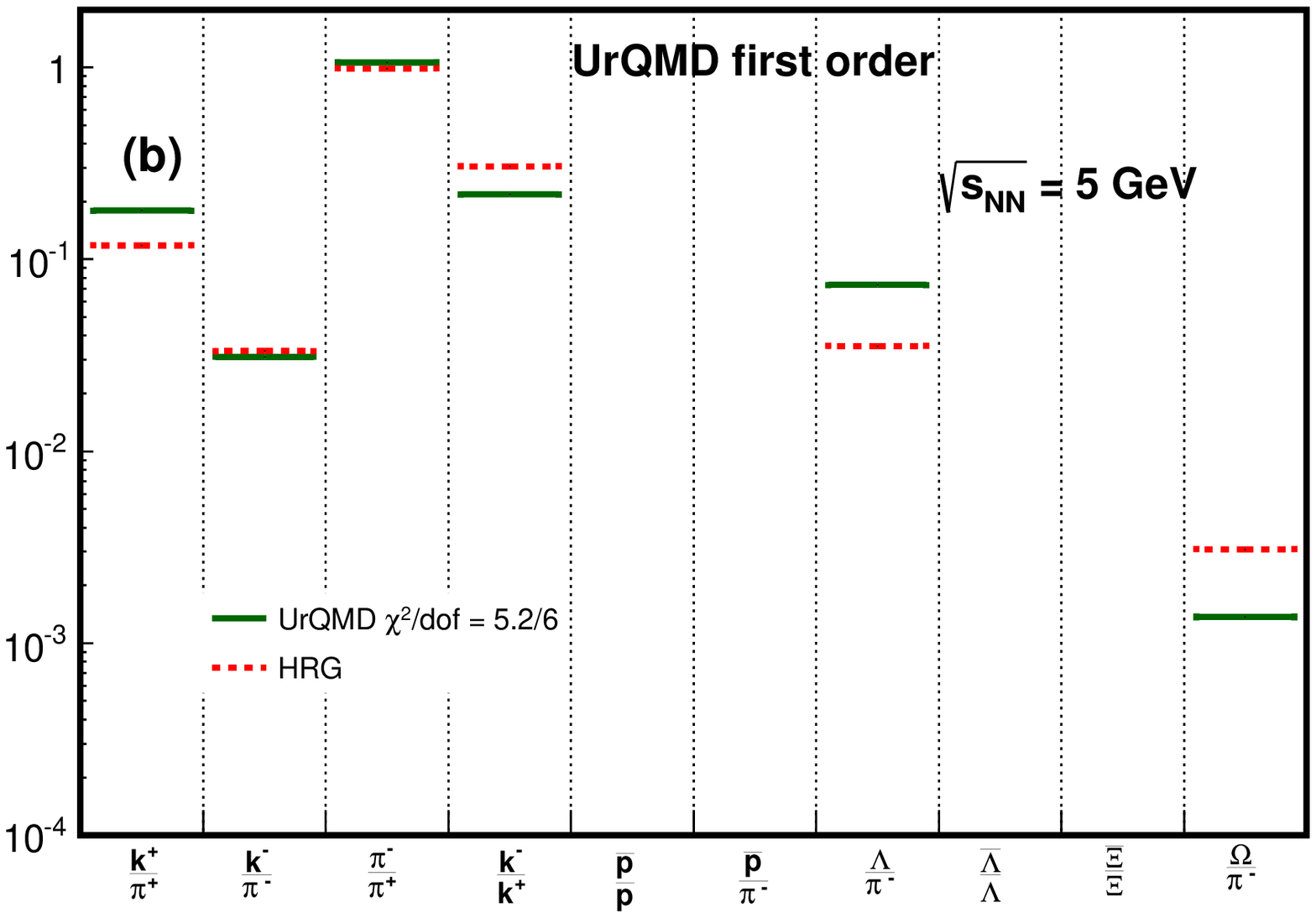}
\includegraphics[width=6.5cm,angle=0]{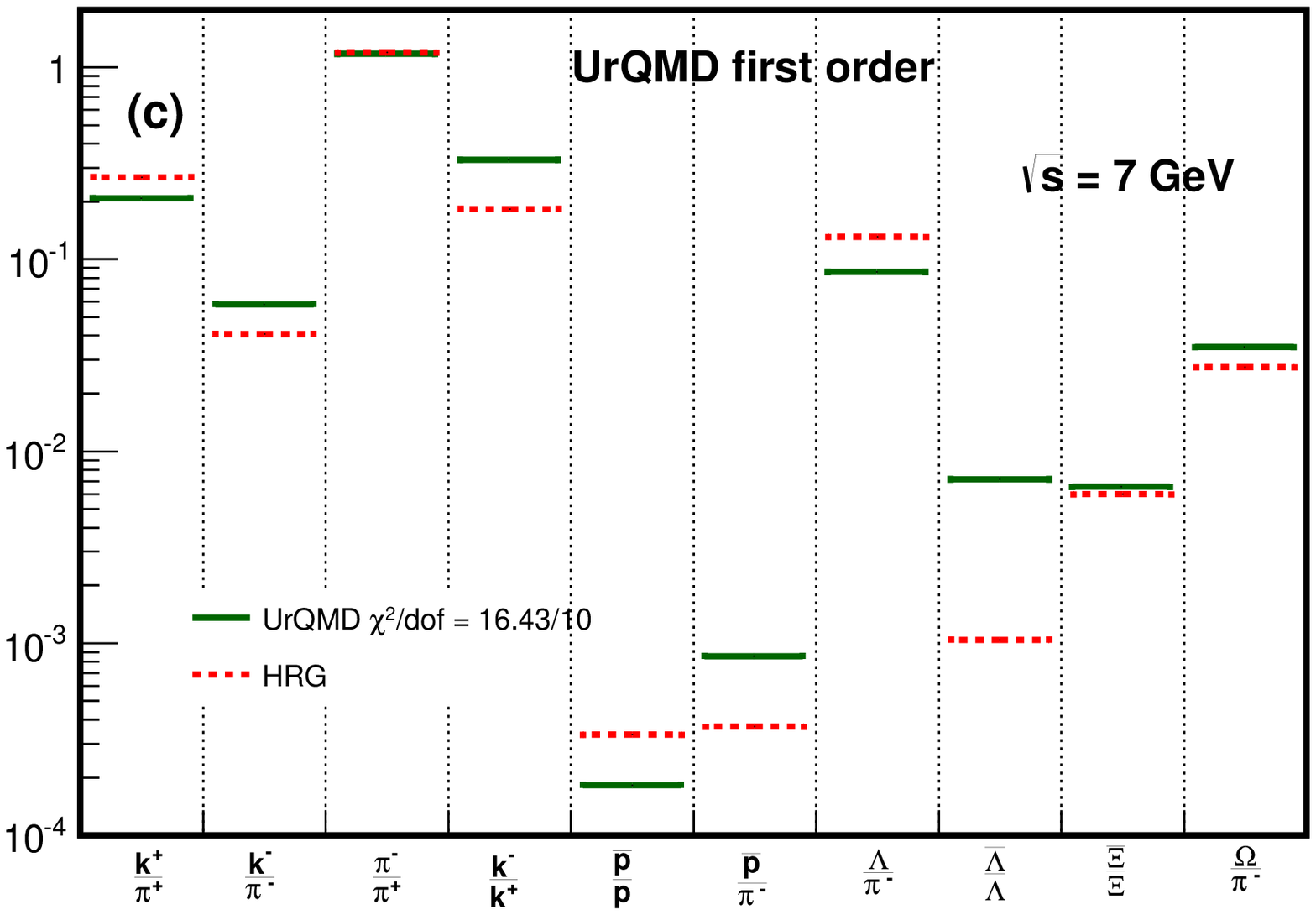}
\includegraphics[width=6.5cm,angle=0]{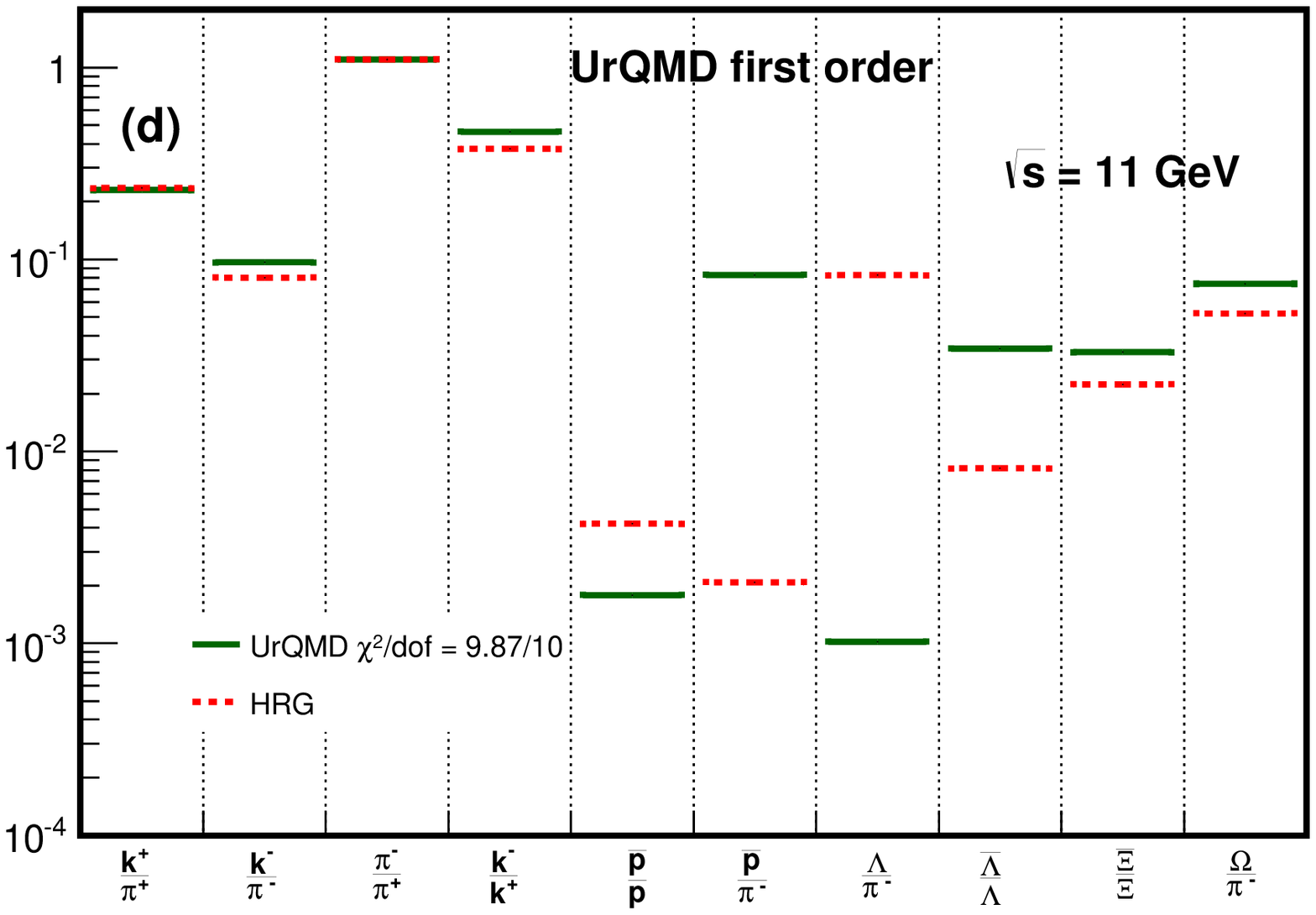}
\includegraphics[width=6.5cm,angle=0]{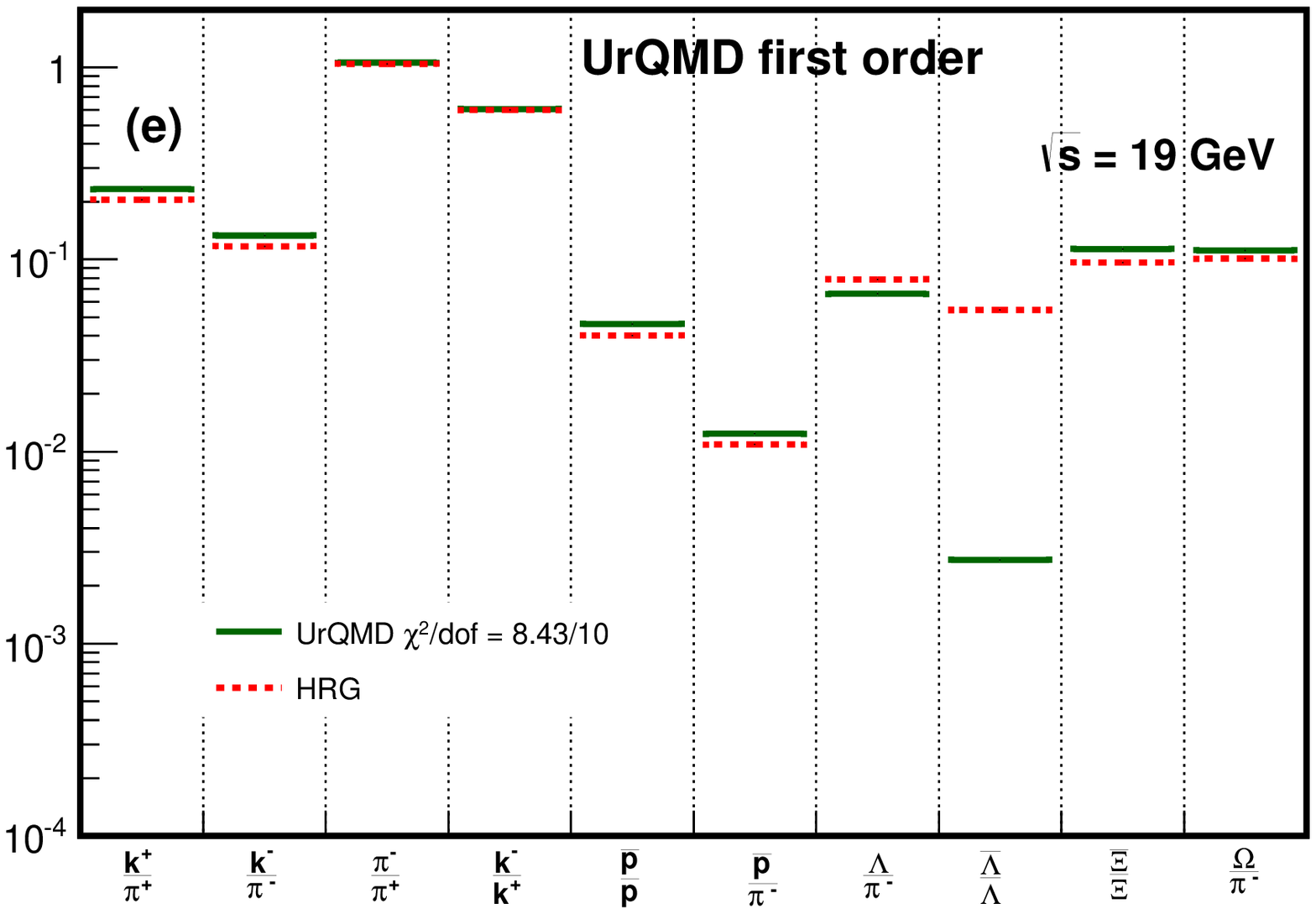}
\caption{The same as in Fig. \ref{fig:crossover}, but here a first-order phase transition is assumed in the hybrid UrQMD simulations at (a) $3$, (b) $5$, (c) $7$, (d) $11$ and (e) $19~$GeV. }
\label{fig:firstorder}
\end{figure}

\begin{figure}[htb]
\includegraphics[width=8cm,angle=0]{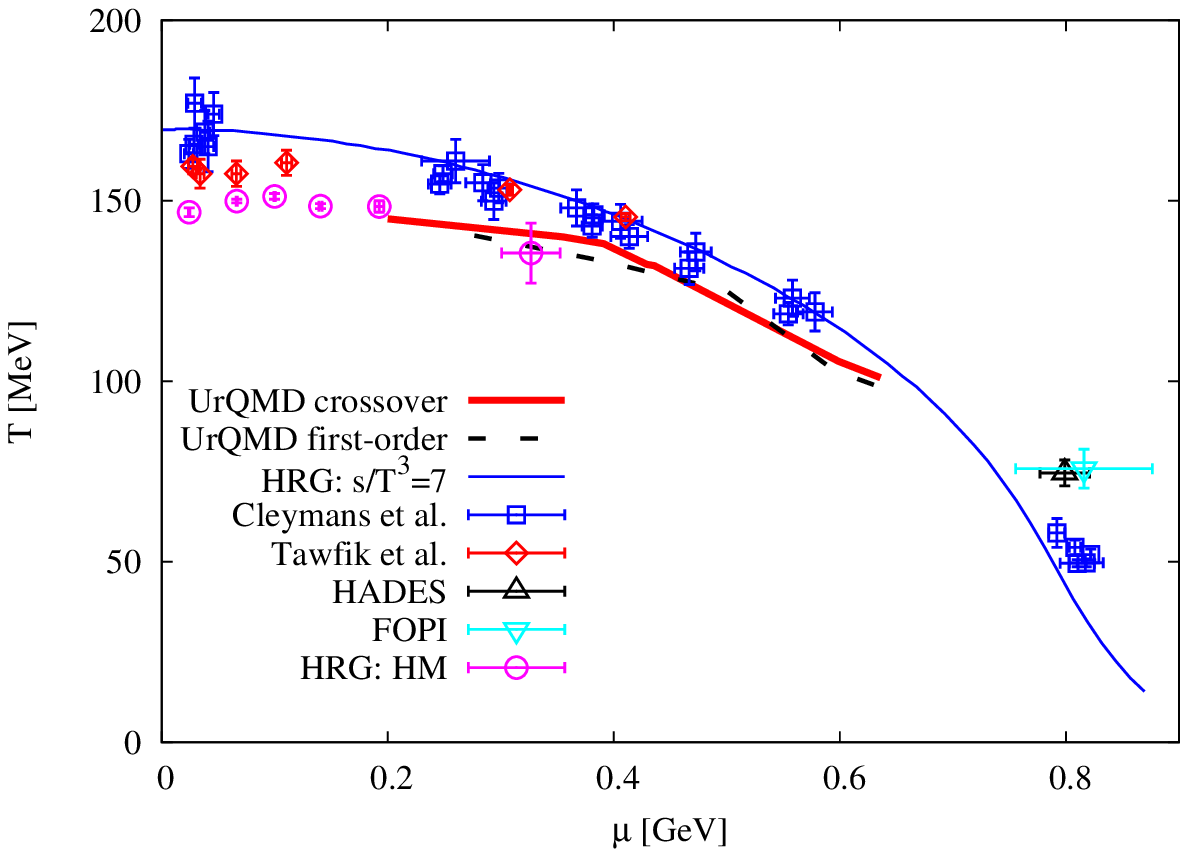}
\caption{The resulting freezeout parameters ($T$ vs. $\mu$) from hybrid UrQMD simulations with  crossover [solid (short) curve] and first-order phase-transition [long-dashed (short) curve] compared with other phenomenological estimations (symbols). }
\label{fig:freezeout}
\end{figure}

Also, for the crossover phase transition, these ten particle ratios are depicted in panels (b) and (c) as well. Other fits for first-order phase transition will be illustrated in Fig. \ref{fig:firstorder}.

The resulting freezeout parameters from the hybrid UrQMD simulations with crossover and first-order phase transition, Tab. \ref{tab:1} and Tab. \ref{tab:2}, respectively, are depicted as thick-solid curve (crossover) and dashed curve (first-order phase transitions) in Fig. \ref{fig:freezeout}. The present calculations are also compared with other estimations (symbols). They are phenomenologically deduced freezeout parameters from measured particle ratios: Cleymans {\it et al.} \cite{clmns}, Tawfik {\it et al.} \cite{Tawfik:2013bza,Tawfik:2014dha}, HADES \cite{hds} and FOPI \cite{fopi} and from measured higher-order moments of net-proton multiplicity: the SU($3$) Polyakov linear-sigma model (PLSM) and HRG \cite{fohm}. The thin curve represents the HRG estimations at the freezeout condition $s/T^3=7$. At a given $\mu_{\mathrm{b}}$ which is related to beam energy $\sqrt{s_{\mathrm{NN}}}$, Eq. (\ref{eq:musqrtsNN}), the freezeout temperature has been determined from the HRG model, section \ref{sec:hrg}, at which the freezeout condition $s/T^3\simeq 7$ is nearly fulfilled.

It is obvious that the UrQMD results agree well with the thermal-model calculations which are based on the higher-order moments of the net-proton multiplicity \cite{fohm}. Also, the calculations from the SU($3$) PLSM are slightly lower that both UrQMD variants. There is a very small difference between UrQMD with crossover and first-order phase transition as can be determined from Tab \ref{tab:1} and Tab. \ref{tab:2}. Accordingly, we conclude that the resulting freezeout parameters are not affected by the type of the phase transition.

The main reason of the small difference between crossover and first-order transitions could be that in the our UrQMD simulations the same value for the particlization criterion  were used in both cases. So, influence of this criterion will be the subject for the next investigations.

When we determine the freezeout parameters from the statistical fits of the HRG calculations to the {\it measured} particle ratios and when we compare them with the fits to UrQMD, we observe that the first ones are relatively higher. This might be due to the assumptions that the constituents of the HRG model are point-like, i.e., no excluded-volume corrections were taken into account, and the quark occupation factors are unity, i.e., light and strange quarks equilibrium, i.e., $\gamma_f$ factors, where $f$ runs over the quark flavors, which are multiplied by the quark occupation parameters $\lambda_i$ in Eq. (\ref{GrindEQ2}). At equilibrium, they are omitted as their values are unity. In nonequilibrium, the quark occupation factors should be stated.

\section{Conclusion}
\label{conc}

Ten particle ratios are generated from the hybrid UrQMD v.$3.4$ at different nucleon-nucleon center-of-mass energies. Two types of the quark-hadron phase-transition; crossover and first-order, are taken into consideration. The energy-dependence of the resulting particle ratios is compared with the HRG calculations and different measured results from the STAR experiments and from the UrQMD model. Within the energy-range considered in this study, a good agreement is observed, at least qualitatively.

We observe that almost all particle ratios from both types of phase transition are nearly indistinguishable, especially at lower energies (larger baryon chemical potentials), which might be interpreted in such a way that the chemical freezeout, at which the particle number should be fixed, apparently takes place immediately after the hadronization process and accordingly the particle production at this chemical equilibrium stage does not differ with respect to its origin. Concretely, we find that for some particle ratios, the simulations with crossover phase transition result in slightly higher temperatures if crossover phase transition is considered than the ones in first-order and vice versa in other ratios. All particle-to-antiparticle ratios are regularly resulting in slightly higher temperatures for crossover phase transition. For these ratios, the agreement between UrQMD or HRG calculations and their measurements is fairly good.

From the energy dependence of the UrQMD particle ratios and the conclusion that the HRG model qualitatively reproduces them and the STAR measurements, as well, we have deduced both freezeout parameters. In doing this, we assume that the UrQMD simulations are experimental inputs. The corresponding uncertainty is determined by statistical errors. We have determined the freezeout parameters at STAR BES; $7.7$, $11.5$ and $19.6~$GeV, whose particle ratios are found compatible with the UrQMD simulations with the crossover phase transition. The resulting freezeout parameters agree well with the ones determined from the statistical-thermal fits of STAR particle ratios at these given energies.

It is found that the resulting freezeout parameters from hybrid UrQMD  agree well with the HRG calculations, in which higher-order moments of the net-proton multiplicity are utilized. Furthermore, the freezeout temperatures deduced from the SU($3$) PLSM are slightly lower that the ones from both of them. We  conclude that the resulting freezeout parameters are not influenced by the order of the quark-hadron phase transition -- or that the aforementioned particlization bias has a possible small influence.

The HRG freezeout parameters determined from the statistical fit of  the {\it measured} particle ratios are relatively higher. This might be understood due to the assumption of point-like constituents and equilibrium light- and strange-quarks occupation factors assumed in the HRG model.

Furthermore, the Parton-Hadron-String Dynamics (PHSD) \cite{PHSDref} and the Three-Fluid Hydrodynamics (3FH) \cite{3FHref} are conjectured as alternatives to perform much better than UrQMD at low energies, towards the NICA energy range. In a future study, we plan to compare between all these approaches at NICA energies.

\section*{Acknowledgments}

The authors thank Prof. V. D. Kekelidze for the fruitful discussions.  The authors also would like to thank the anonymous referee for the important comments and suggestions to improve the manuscript. The work was supported  by the project "Membership in MPD and BM@N collaborations at NICA" within  the JINR-Egypt programme. 



\end{document}